\documentclass[preprint,11pt]{elsarticle}
\usepackage[left=2.5cm,right=2.5cm,top=2cm,bottom=3cm]{geometry}
\usepackage{graphicx} % Required for inserting images
\usepackage{siunitx}
\usepackage{placeins}
\usepackage{arydshln}
\usepackage{diagbox}

\usepackage{hyperref}
\hypersetup{
    colorlinks=true,
    linkcolor=blue,
    filecolor=magenta,      
    urlcolor=cyan,
    citecolor=blue,
}

\newcommand{\eg}{e.g.\ }
\newcommand{\ie}{i.e.\ }
\newcommand{\Rsq}{$\text{R}^2$}
\newcommand{\wera}{w_\text{ERA-5}}
\newcommand{\wclim}{w_\text{clim}}
\newcommand{\wraw}{w_\text{raw}}
\newcommand{\wcorr}{w_\text{corr}}

\newcolumntype{C}[1]{>{\centering\let\newline\\\arraybackslash\hspace{0pt}}m{#1}}

\usepackage{natbib}
\setcitestyle{authoryear,open={(},close={)},citesep={;}}

\journal{Applied Energy}

\begin{document}

\begin{frontmatter}

\date{}

\title{The future of offshore wind power production: wake and climate impacts}

\author[inst1]{Simon C. Warder}
\author[inst1]{Matthew D. Piggott}

\affiliation[inst1]{organization={Department of Earth Science and Engineering},
            addressline={Imperial College London}, 
            city={London},
            postcode={SW7 2BP}, 
            country={UK}}

\begin{abstract}
Rapid deployment of offshore wind is expected within the coming decades to help meet climate goals. With offshore wind turbine lifetimes of 25--30 years, and new offshore leases spanning 60 years, it is vital to consider long-term changes in potential wind power resource at the farm planning stage. Such changes may arise from multiple sources, including climate change, and increasing wake-induced power losses. In this work, we investigate and compare these two sources of long-term change in wind power, for a case study consisting of 21 wind farms within the German Bight. Consistent with previous studies, we find a small but significant reduction in wind resource due to climate change by the end of the 21st century under the high-emission RCP8.5 scenario, compared with a historical period, with a mean power reduction (over an ensemble of seven climate models) of 2.1\%. To assess the impact of wake-induced losses due to increasingly dense farm build-out, we model wakes within the German Bight region using an engineering wake model, under various stages of (planned) build-out corresponding to the years 2010--2027. By identifying clusters of wind farms, we decompose wake effects into long-range (inter-cluster), medium-range (intra-cluster) and short-range (intra-farm) effects. Inter-cluster wake-induced losses increase from 0 for the 2010 scenario to 2.5\% for the 2027 scenario, with intra-cluster losses also increasing from 0 to 4.3\%. Intra-farm losses are relatively constant, at around 13\%. While the evolution of wake effects therefore outweighs the climate effect, and impacts over a shorter timescale, both factors are significant. We also find evidence of an interaction between the climate and wake effects. Both climate change and evolving wake effects must therefore be considered within resource assessment and wind farm planning.
\end{abstract}


\end{frontmatter}

%\maketitle

%\FloatBarrier

%\linenumbers
\setcounter{page}{1}

\section{Introduction}

Future deployment of offshore wind farms is a key component of plans to meet climate goals, with rapid growth expected over the coming decades \citep{gwec_2023}. The North Sea, due to its high wind resource and shallow seas, is particularly well suited to offshore wind deployment, and currently accounts for more than 70\% of European offshore wind capacity \citep{wind_farm_data,gwec_2023}. Regions such as the German Bight have seen particularly rapid growth; see table \ref{tab:farms}. With wind farm lifetimes typically spanning 20--25 years \citep{topham2017sustainable,ziegler2018lifetime}, and new offshore leases in the UK lasting 60 years \citep{crown_estate_round_4}, it is vital to consider long-term changes in the wind resource at the wind farm planning stage. In this work, we focus on two significant factors which may influence the power output from offshore wind farms on these timescales. These are (i) wake-induced effects due to the increasing density of wind farm build-out, and (ii) long-term changes in the underlying wind resource, due to climate change. We review the literature in these areas in the following subsections.

\subsection{Wake effects}
\label{sec:intro_wakes}

Wind power output is impacted by wake effects across a variety of spatial scales \citep{pryor2024power}, and can reduce wind farm power output by 10--40\% \citep{barthelmie2009modelling,fei2020experimental,wang2024wind}. Wakes arise due to the extraction of kinetic energy by wind turbines, resulting in reduced wind speeds and increased turbulence intensity downstream \citep{goccmen2016wind}. This in turn reduces the total power which can be extracted by a wind farm and also increases the fatigue experienced by wind turbines. At larger scale, wind farm wakes arising from the accumulation of individual turbine wakes impact on the power produced by other farms downstream \citep{pryor2024wind}.

Intra-farm wakes refer to individual turbine wakes, which impact on other turbines within the same farm. Intra-farm wakes are well studied in the literature. A typical intra-farm problem is that of array layout optimisation to maximise power output or minimise LCoE \citep{azlan2021review,kumar2021layout}. For example, \citet{thomas2023comparison} compared optimisation methods for the Borssele III and IV farms in the North Sea, using a Gaussian wake model, achieving reductions in wake losses from 17.28\% to between 15.48 and 15.70\% depending on the optimisation method. \citet{sickler2023offshore} studied the effect of regular and irregular wind farm layouts, finding the irregular layout to increase power by 0.66\% for a 74-turbine real-world case study, but also increasing fatigue by up to 24\%.

Longer-range wakes are often referred to in the literature as inter-array \citep{pryor2024power}, inter-farm \citep{feroz2020wind}, farm-to-farm \citep{fischereit2022comparing} or cluster wakes \citep{nygaard2020modelling}, and their impact is known as wind shadowing \citep{schneemann2020cluster,finseraas2024gone} or wind theft \citep{pryor2020wind}. Such effects have gained significant attention only over the last few years. These long-range wakes can propagate 10s of \si{\kilo\metre} downstream \citep{platis2018first,schneemann2020cluster,canadillas2020offshore,li2023long}, depending on atmospheric conditions \citep{abkar2015influence,foreman2024atmospheric,rosencrans2024seasonal}. \citet{vimalakanthan2024large} reveal measurement data indicating wake deficits in excess of 5\% up to 43 \si{\kilo\metre} downstream. For a future US East Coast scenario, \citet{pryor2024wind} find that wind farms will produce wakes covering 25\% of available unleased area. \citet{rosencrans2024seasonal} also consider a future large-scale US East Coast scenario, differentiating between intra- and inter-farm losses, with large contributions from both internal wakes (25.7--29.2\%) and external wakes (13.4--14.7\%).

A number of wake modelling approaches exist, with different methods suited to different scales and applications \citep{amiri2024review,wang2024wind}. CFD models \citep{sanderse2011review} can provide high-fidelity simulations of wakes from individual turbines or within farms, but are computationally expensive and therefore not suited to applications requiring a large number of simulations, such as array layout optimisation, or to the large spatial domains required to model inter-farm effects. Numerical weather prediction (NWP) models are also capable of simulating wake effects, but the spatial resolution at which they are typically applied is best suited to inter-farm/cluster wake analysis \citep{pryor2024wind}, while their computational cost is also too high for use within optimisation studies.
So-called engineering wake models have been developed since the 1980s \citep{jensen1983note,katic1986simple}, and remain a commonly-used approach for simulating the effect of wakes on turbine power output. Although engineering models may use a variety of methods including analytic approaches or lookup tables, they are characterised by their strong simplifying assumptions (\eg spatially and temporally uniform flow) but low computational cost, and are thus commonly applied to layout optimisation and control problems \citep{azlan2021review}. On these scales, the simplifying assumptions hold fairly well and the low computational cost can be leveraged for efficient optimisation.

In particular, the TurbOPark engineering wake model \citep{nygaard2020modelling,pedersen2022turbulence} has gained popularity, having been developed in an attempt to accurately capture wake decay over long (\eg inter-farm/cluster) distances. The overestimation of wake recovery was a known issue with earlier engineering wake models \citep{stieren2021evaluating,fischereit2022comparing}, resulting in underestimation of inter-farm wake effects \citep{pedersen2022turbulence}. TurbOPark has been previously validated using data from observations and high-fidelity CFD modelling. For example, \citet{pedersen2022turbulence} and \citet{nygaard2022large} benchmarked the model against SCADA data in comparison with the earlier `Park' model, demonstrating the superior performance of the TurbOPark model. \citet{van2024performance} compared several calibrated engineering wake models against SCADA data from the Belgian-Dutch offshore zone, finding that TurbOPark performs best among the engineering wake models tested, although all engineering models shared similar trends once calibrated. \citet{sorensen2024extension} also compared TurbOPark against power production data, for a case study comprising the Lillgrund, Rødsand and Horns Rev farms in the North Sea, finding a mean absolute percentage error of 4.1\%, which was the best performance among the engineering wake models they tested. \citet{zum2024evaluation} compared in situ airborne measurements against model results using both an NWP model, and TurbOPark. Both modelling approaches compared well with the observations up to 20--30\si{\kilo\metre} downstream of the wind farm, although the performance of the NWP approach remained more consistent further downstream. Finally, \citet{stieren2021evaluating} benchmarked several engineering wake models against high-fidelity LES simulations of a pair of wind farms. While all considered engineering models overpredicted the wake recovery compared to the LES simulations, the TurbOPark model performed best. Overall, these studies indicate that TurbOPark performs well in capturing long-range interactions up to scales of 10s of \si{\kilo\metre}, and is widely acknowledged to be the most accurate of the currently available engineering wake models. It is for this reason, along with its computational efficiency, that we choose to employ the TurbOPark model within this study.

Wake effects, particularly on the inter-farm scale, evolve on annual to decadal timescales due to the rapid expansion of offshore wind. In addition to \citet{pryor2024wind} and \citet{rosencrans2024seasonal}, both discussed above, \citet{pryor2024power} also model future wind farms along the US east coast, predicting wake losses of up to 37\% depending on turbine deployment scenario and wake parameterisation method. \citet{akhtar2021accelerating} studied wake impacts due to all currently planned farms in the North Sea using a mesoscale NWP approach. They found wakes extending 40 \si{\kilo\metre} downstream during prevailing wind conditions, resulting in capacity factor reductions in excess of 20\%. \citet{van2023simulating} model a future wind farm cluster in the North Sea with a total capacity of 10 \si{\giga\watt}, finding wake-induced losses ranging from 5.7--10.1\% depending on the modelling approach, of which 0.46--3.6\% was due to inter-farm interactions. \citet{borgers2023mesoscale} consider a future wind farm scenario in the North Sea based on current and planned wind farms, including an investigation of the influence of turbine type on cluster wake losses.

Although some of the above studies include sensitivity analysis with respect to the types and densities of the turbines deployed, they still consist of only a single build-out stage, and do not focus on the time evolution of build-out. This time-evolution is central to long-term policy and decision making, and contributes to current regulatory gaps regarding the impact of inter-farm wakes \citep{lundquist2019costs,finseraas2024gone}. It is for these reasons that the present study includes an analysis of the time evolution of wake effects due to build-out, over a period of two decades. Furthermore, while existing studies typically focus on either intra-farm or inter-farm/cluster wake effects, here we distinguish between effects at different spatial scales, in order to gain insight into the development of wake losses with build-out.

\subsection{Climate change}
\label{sec:intro_climate}

Climate change is expected to impact future wind power output. The primary mechanism for this is via changes in mean wind speed, although contributions may also arise from changes in the distribution of winds, temperature, sea level rise or extreme weather events \citep{solaun2019climate}. Although there are significant uncertainties in predicting future changes in wind power due to climate change, climate projection models are considered the most well-trusted approach \citep{solaun2019climate}, with the Coupled Model Intercomparison Project phase 5 (CMIP5) \citep{taylor2012overview} and phase 6 (CMIP6) \citep{eyring2016overview} providing ensembles of global climate projections under a variety of scenarios.

In the North Sea, including the German Bight which is the focus of this paper, ensembles from CMIP5 \citep{carvalho2017potential} and EURO-CORDEX \citep{jacob2014euro} (which is downscaled from CMIP5) \citep{moemken2018future} indicate a decrease in mean wind resources, but an increase in variability \citep{susini2022climate}. Based on CMIP6 projections for a medium-term horizon (2031--2050), \citet{hahmann2022current} found that mean wind resource in the North Sea is not particularly affected by climate change, although the seasonal distribution is ``significantly altered'', with reduced wind in summer found across much of the North and Baltic seas. Comparing CMIP5 and CMIP6, \citet{carvalho2021wind} show that the magnitude and spatial distribution of future changes in wind resource depend on the emissions pathway; under the SSP5-8.5 scenario, resources decrease for most of Europe by the end of the century, while for SSP2-4.5 there is a small increase in resources for parts of the domain (Ukraine and Turkey), while all other regions decrease. Their results show high levels of uncertainty in terms of changes in variability. \citet{martinez2023evolution} employ CMIP6 to study changes in offshore wind resource in Northern Europe. Consistent with the findings of the other studies cited above, they find declining wind resources across much of Europe by the end of the 21st century, with the exception of parts of the Baltic Sea, which experience a small increase.

Considering elsewhere in Europe, \citet{fernandez2023dynamic} study changes in wind speed over the North Atlantic Ocean, using downscaled CMIP6 projections. Their results indicate significant summer and autumn increases in wind power density off the Iberian Peninsula, for both mid- and late- century time periods, under a variety of emissions pathways, resulting in higher stability in offshore resource across the year. This is in contrast to the North Sea, for which their results indicate a decrease in resources, consistent with the other studies above. The increase in resources for the Iberian Peninsula is consistent with the results of \citet{thomas2023downscaling}, who also utilise downscaled CMIP6 projections. \citet{abdelaziz2023long} assess the impact of future climate change in UK waters, finding significant decreases in wind resource in summer, and identifying the north and west regions of UK waters being the most resilient to climate change, while the east (where the majority of existing farms are sited) suffers stronger adverse effects. Beyond Europe, a large number of studies have investigated global climate impacts on wind resources, which are consistent with the regional-scale studies above in terms of decreasing resources across northern Europe \citep[e.g.][]{martinez2024global}.

Common to many of these studies of climate impact on wind resources is the need for bias correction. For example, wind speeds from reanalysis products may be bias-corrected to produce national power outputs consistent with production data \citep{staffell2016using}. In the climate context, bias correction is considered vital for wind energy applications, since climate models are not specifically designed for projecting wind resources \citep{hdidouan2017impact}. \citet{luzia2023validating} employed data from the Global Wind Atlas (GWA) to bias-correct EURO-CORDEX downscaled climate simulations. This took advantage of the relatively high spatial resolution of GWA, and resulted in a significant reduction in the model ensemble spread. \citet{moemken2018future} employed ERA-interim reanalysis to bias-correct EURO-CORDEX simulations over Europe, based on a statistical approach employing fitted Weibull distributions, in order to increase confidence in projected changes in wind resources over Europe. \citet{li2019statistical} also utilised reanalysis data to bias-correct CORDEX simulations, although they used only a single climate model, with a case study region in East Asia. They compared several statistical bias correction methods including linear scaling, variance scaling and quantile mapping, concluding that the best performance was obtained using quantile mapping based on fitted Weibull distributions. \citet{costoya2020using} employed a combination of quantile mapping and correction in the frequency domain to CORDEX simulations, resulting in reduced error in modelled wind power density in the Iberian Peninsula. \citet{shen2024increases} applied a PCA-based bias correction to CMIP6 projections using ERA-5 as `ground truth', which they found helped to address discrepancies in long-term trends.

While the impact of climate change on wind energy resources in the North Sea is reasonably well studied, including the need for bias correction, such studies typically neglect wind turbine and farm wakes. An exception is \citet{hahmann2022current}, who investigated the impact of climate change on wind energy resource in the North Sea, utilising an ensemble of CMIP6 projections. The study included wind turbine wake effects based on a planned Energy Island cluster of wind farms, modelled using a simple engineering wake model \citep{jensen1983note}. They found a significant decrease in energy production during summer of 6.9\% for the 2031--2050 period under the SSP5-8.5 scenario, of which 0.7\% was due to wake effects. These results suggest that the interaction between climate and wake effects warrants further investigation. Furthermore, any long-term studies of wind resource, spanning farm lifetimes or offshore lease lengths, may suffer from systematic errors unless both climate and wake effects are considered.

\subsection{Contributions of this study}
\label{sec:intro_contribs}

As described in section \ref{sec:intro_wakes}, there have been limited studies into the time-evolution of wake effects, or studies which break down wake effects by spatial scales. These aspects are particularly important in regions such as the German Bight, where current wind farm installations are already densely packed, and with further developments planned \citep{wind_farm_data}.

Long-term changes in wind resources due to climate change are, in contrast, very well studied. However, the findings of \citet{hahmann2022current} described above suggest that there is potential benefit to further studies into the interaction between climate change and wind turbine wake effects, and that omitting wake effects when assessing climate change may lead to misleading conclusions.

The present study addresses these gaps. Specifically, the novel contributions of this work are
\begin{enumerate}[(i)]
    \item a study of the time-evolution of wake effects over two decades, within an increasingly densely packed area of the North Sea;
    \item the break-down of these wake losses into intra-farm (short-range), intra-cluster (medium-range) and inter-cluster (long-range) effects;
    \item a comparison of climate and wake impacts, which are found to both have a significant impact over the time scale of wind farm lease durations;
    \item an investigation into the interaction of climate and wake effects, or equivalently how wake impacts change in future with climate change.
\end{enumerate}

The remainder of the paper is structured as follows. The methods are presented in section \ref{sec:methods}, including a description of the study region, wind farm data, reanalysis and climate projection datasets, bias correction methods, and wake modelling approach, as well as a description of the experiments we perform. The corresponding results are presented in section \ref{sec:results}. A discussion of the results, including the limitations of this study and avenues for further work, can be found in section \ref{sec:discussion}. Finally, we draw conclusions in section \ref{sec:conclusions}.

\section{Methods}
\label{sec:methods}

\subsection{Study region \& wind farm data}
\label{sec:methods_study_region_and_farm_data}

\begin{figure}
    \centering
    \includegraphics[width=0.7\textwidth]{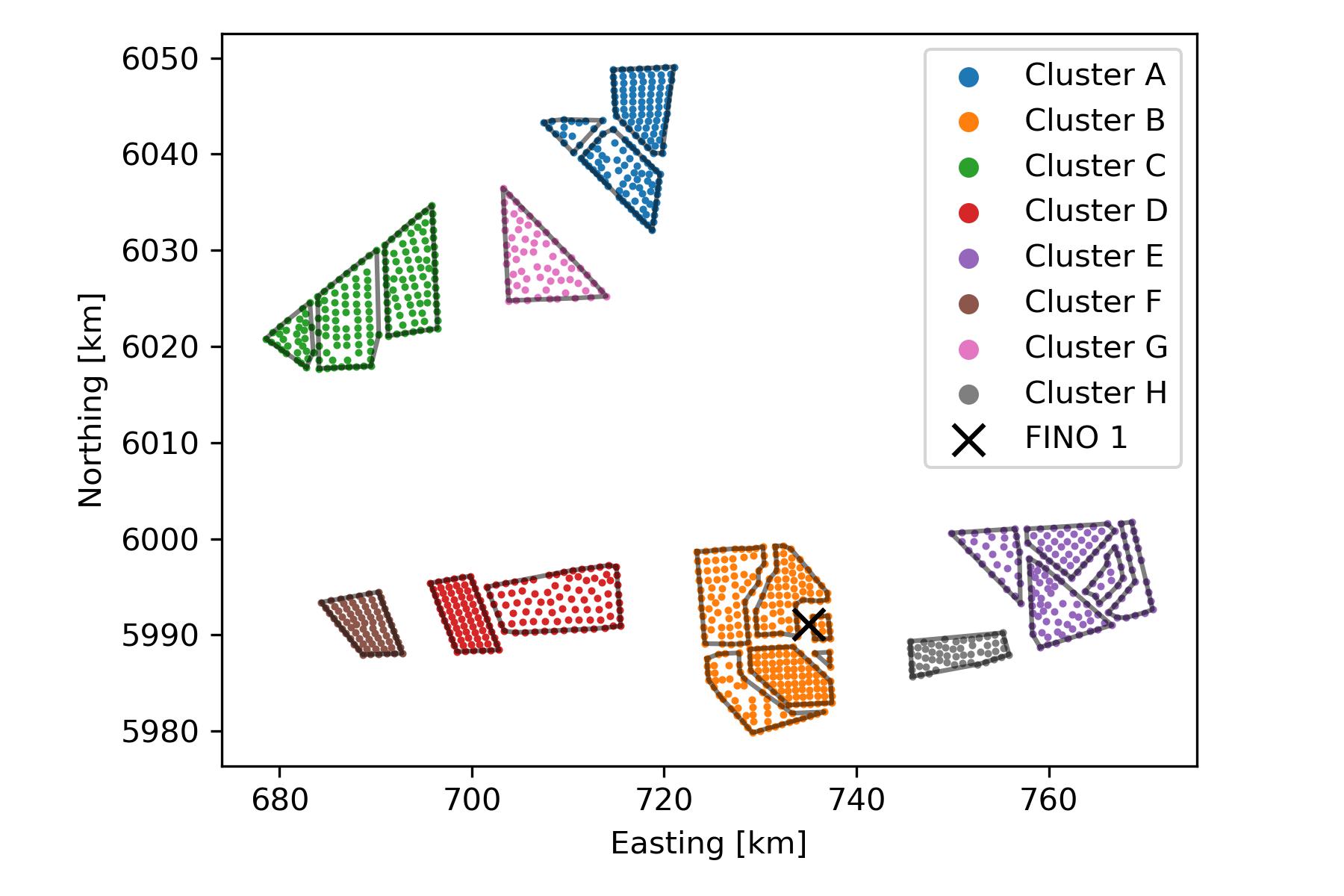}
    \caption{Wind farms used within this study. The outlines of each farm are in grey, with the turbine colour indicating the cluster ID, as given by the legend. For further detail on each wind farm see table \ref{tab:farms}. We use data from the FINO1 platform for the purposes of model validation. The platform is situated within cluster B, adjacent to the Alpha Ventus farm.}
    \label{fig:farms}
\end{figure}

As a case study region, we choose a set of wind farms in the German Bight region of the North Sea. Our choice of case study region was motivated by two factors. Firstly, a large number of wind farms have been built in the region over a number of years, as summarised in table \ref{tab:farms} and shown in figure \ref{fig:farms}. Secondly, the FINO1 observation platform, also indicated in figure \ref{fig:farms}, predates any of the wind farms and constitutes a valuable source of validation data. The spatial extent of our case study region is selected to be sufficiently large that simulations of power output are impacted by wind shadowing effects at the inter-cluster scale, but not so large that the simulations we perform become too computationally expensive. The next nearest wind farm to those included in the case study is 45 \si{\kilo\metre} away.

\begin{figure}
    \centering
    \includegraphics[width=0.3\linewidth]{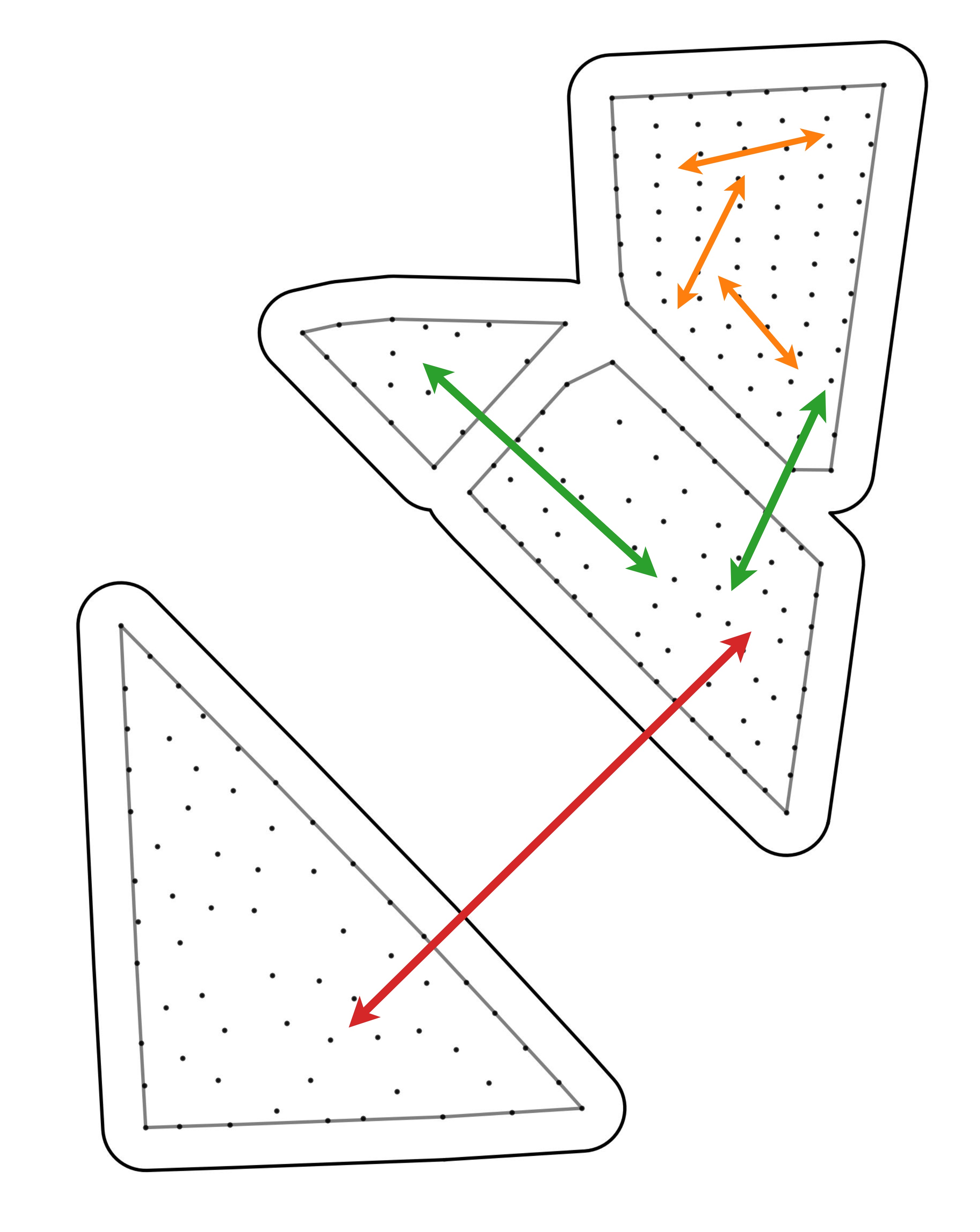}
    \caption{Schematic diagram of intra-farm (orange), intra-cluster (green) and inter-cluster (red) interactions.}
    \label{fig:wake_scales}
\end{figure}

Wind turbine locations are from \citet{zhang2021global}, which we supplement using data for planned future farms from \citet{wind_farm_data}, including only those farms with known turbine models and layouts. We note that there are a number of further wind farm developments planned within our case study region; the inclusion of these farms is left to future work, since at present there is limited publicly available detail about them. We group the farms into small clusters as indicated in figure \ref{fig:farms}, which we use as the basis for distinguishing between short (intra-farm), medium (intra-cluster) and long (inter-cluster) distance wake effects; see figure \ref{fig:wake_scales}.

Most wind turbine power curves are not publicly available. We instead generate plausible power curves using PyWake's `generic\_wind\_turbines' module \citep{mads_m_pedersen_2019_2562662}, based on publicly available turbine data \citep{wind_turbine_models} (nominal power, hub height, rotor diameter, cut-in and cut-out speeds). This requires further assumptions regarding mechanical and electrical losses, as well as the peak thrust coefficient. We find that the peak thrust coefficient is particularly influential (and uncertain), and we therefore include a sensitivity analysis with respect to this parameter in \ref{sec:sensitivity_to_ct}.

\begin{table}[]
    \centering
    \begin{tabular}{c|c|c|c|c}
        Farm name & Commission date & Installed capacity / \si{\mega\watt} & No. turbines & Cluster ID \\ \hline
        Global Tech 1 & 2015-09-02 & 400 & 80 & A \\
        Hohe See & 2019-11-01 & 497 & 71 & A \\
        Albatros & 2020-01-09 & 112 & 16 & A \\ \hdashline
        Alpha Ventus & 2010-04-27 & 60 & 12 & B \\
        Borkum Riffgrund 1 & 2015-10-09 & 312 & 78 & B \\
        Borkum Riffgrund 2 & 2018-12-31 & 448 & 56 & B \\
        Merkur & 2019-06-24 & 396 & 66 & B \\
        Trianel Borkum II & 2020-07-03 & 360 & 72 & B \\ \hdashline
        Bard Offshore 1 & 2013-08-26 & 400 & 80 & C \\
        Veja Mate & 2017-05-31 & 402 & 67 & C \\
        Deutsche Bucht & 2019-09-30 & 248 & 31 & C \\ \hdashline
        Gemini (east) & 2017-04-28 & 300 & 75 & D \\
        Borkum Riffgrund 3 & 2025-12-31 & 913 & 83 & D \\ \hdashline
        Gode Wind 1 & 2016-09-30 & 330 & 55 & E \\
        Gode Wind 2 & 2016-09-30 & 252 & 42 & E \\
        Gode Wind 3 & 2024-12-31 & 253 & 23 & E \\ 
        Nordsee Cluster A N-3.7 & 2027-01-01 & 240 & 16 & E \\
        Nordsee Cluster A N-3.8 & 2027-01-01 & 450 & 30 & E \\ \hdashline
        Gemini (west) & 2017-04-28 & 300 & 75 & F \\ \hdashline
        EnBW He Dreiht & 2025-12-31 & 960 & 64 & G \\ \hdashline
        Nordsee 1 & 2017-12-20 & 334.8 & 54 & H \\
    \end{tabular}
    \caption{Wind farms located within our study region. There are 21 farms, with a total of 1,146 turbines, and a combined installed capacity of 7,967.8 \si{\mega\watt}.}
    \label{tab:farms}
\end{table}

%\FloatBarrier

\subsection{Reanalysis data}

Reanalysis datasets are derived from historical weather simulations, into which a variety of observation data has been assimilated. They therefore represent a trusted source of historical weather data as a gridded product. In this work we use the ERA-5 global reanalysis dataset \citep{hersbach2020era5}. ERA-5 has a spatial resolution of 0.25\si{\degree}, and a temporal resolution of one hour, covering the period from 1940 to present. From this, we use the 100 \si{\metre} wind velocity components, at 3-hourly temporal resolution. 

We use ERA-5 data for two purposes within this work. The first is to provide the base atmospheric state within which we validate our wake modelling workflow against observations. Since `historical' climate model data are not synchronised with the observed weather, we cannot make direct comparisons between climate datasets and observations, but instead rely on comparisons of long-term statistics. We are therefore able to more robustly validate our wake modelling workflow by applying it to reanalysis data and directly comparing with observational data from FINO1. For this purpose, we use ERA-5 data from the period 2004--2023.

The second use of ERA-5 reanalysis data within this work is as the basis for bias-correcting the climate datasets, for which we use the period 1976--2006. This choice is motivated by several factors. Firstly, ERA-5 has global spatial coverage, which will enable the application of our methodology to new locations in future work. Secondly, it covers a long historical period, which is crucial for making statistical comparisons with historical climate projections, which are not synchronised with historical weather. Finally, the high temporal resolution of the available data enables flexibility in terms of the bias correction methods applied, in contrast to wind atlas products which may only provide certain wind speed statistics.

\subsection{Climate projection data}

In this study we utilise regional climate simulations from EURO-CORDEX \citep{jacob2014euro}, accessed via \citet{esgf}. These simulations are downscaled from CMIP5 climate models \citep{taylor2012overview}, on a regular 0.11$^\circ$ resolution grid (or approximately 12.5 \si{\kilo\metre}). Although CMIP5 has been superseded by CMIP6, the availability of downscaled CMIP6 projections remains limited, hence we use EURO-CORDEX projections. The specific EURO-CORDEX models we choose are summarised in table \ref{tab:euro-cordex_models}. These models were selected on the basis that they provide wind speed data at 100 \si{\metre} height, at 3-hourly temporal resolution. The same set of models have previously been evaluated for the purposes of wind power generation by \citet{luzia2023validating}.

The time periods we consider are 1976--2006 and 2070--2100, and a further period of 2040--2070 is also analysed in \ref{sec:alt_climates}. The first period is derived from `historical' EURO-CORDEX simulations, and for the latter two periods we use data corresponding to the high-emission RCP8.5 scenario. This scenario is the most commonly studied in the literature \citep{jung2022review}, but results should be interpreted in the context that it is expected to produce the most significant climate change signal. Our choice to use the RCP8.5 scenario is based on the availability of downscaled climate model outputs at our desired spatial and temporal resolutions. In \ref{sec:alt_climates} we also present results from the RCP2.6 scenario, for the 2070--2100 period.

\begin{table}[]
    \centering
    \footnotesize
    \begin{tabular}{c|c|c|c}
        Forcing GCM/ESM & RCM & Abbreviation & References \\ \hline
        CNRM-CERFACS-CNRM-CM5 & ALADIN63 & CNRM-A & \citet{voldoire2013cnrm,nabat2020modulation}\\
        NCC-NorESM1-M & ALADIN63 & NCC-A & \citet{bentsen2013norwegian} \\
        MOHC-HadGEM2-ES & ALADIN63 & MOHC-A & \citet{bellouin2011hadgem2} \\
        CNRM-CERFACS-CNRM-CM5 & RegCM4–6 & CNRM-R & \citet{coppola2021non} \\
        NCC-NorESM1-M & RegCM4-6 & NCC-R & \citet{ncc_data} \\
        ICHEC-EC-EARTH & RegCM4–6 & ICHEC-R & \citet{hazeleger2012ec} \\
        MPI-M-MPI-ESM-LR & RegCM4–6 & MPI-R & \citet{giorgetta2013climate} \\
    \end{tabular}
    \caption{Summary of EURO-CORDEX simulations used within this study, including the underlying General Circulation Model (GCM) or Earth System Model (ESM) used as forcing for the EURO-CORDEX model, and the Regional Circulation Model (RCM) employed. The abbreviations will be used throughout the rest of this paper to refer to each model.}
    \label{tab:euro-cordex_models}
\end{table}

\subsection{Observation data}

The FINO1 offshore observation platform is located within our study region, as indicated in figure \ref{fig:farms}, and has been operational since 2004, predating all farm developments in the area. Data from FINO1 constitutes a valuable source of ground truth data. In this work, we use wind speed data from a cup anemometer at a height of 102 \si{\metre}, which has been corrected for mast effects (see \citet{westerhellweg2011comparison,westerhellweg2012fino1}). We use this data to validate our reanalysis dataset ERA-5, as well as our wake modelling workflow.

\subsection{Bias correction methods}

As is typical in bias correction for climate applications, we use a historical comparison period to calculate or train the bias correction parameters, with this correction then applied to the future projected dataset \citep{hdidouan2017impact}. This implicitly assumes that the biases in the future climate projections are consistent with those of the historical data. As described above, we use ERA-5 reanalysis as the basis for bias-correcting downscaled climate models. In the following, the (historical) ERA-5 wind speed at a given location is denoted $\wera$, and the corresponding wind speeds from the historical downscaled climate dataset are denoted $\wclim$. The raw and corrected future climate projection data are denoted $\wraw$ and $\wcorr$, respectively.

In this work, we apply and compare five common statistical bias correction methods to the wind speeds predicted by each climate model. This allows us to select the best-performing method, which we subsequently use for quantifying future climate change. Each method is performed on a per-month basis in order to correct for seasonal bias. We first regrid ERA-5 data to align with the climate model grid. We then apply the bias correction for each grid point in turn, resulting in a spatially varying bias correction. We detail each of the bias correction methods in the following subsections.

\subsubsection{Offset}

A simple offset is applied to all wind speeds to align the mean of $\wclim$ with that of $\wera$. The corrected wind speed is given by
\begin{equation}
    \label{eq:add_lin_scaling}
    \wcorr = \wraw + \mu(\wera) - \mu(\wclim),
\end{equation}
where $\mu$ denotes the mean.

\subsubsection{Variance scaling}

Variance scaling is similar to the offset correction above, with the additional step of rescaling based on the standard deviation of the historical datasets. The corrected speed is given by
\begin{equation}
    \label{eq:var_scaling}
    \wcorr = \left( \wraw - \mu(\wraw) \right) \frac{\sigma(\wera)}{\sigma(\wclim)} + \mu(\wraw) + \mu(\wera) - \mu(\wclim),
\end{equation}
where $\sigma$ denotes the standard deviation.

\subsubsection{Multiplicative scaling}

Wind speeds are scaled by a constant multiplicative factor to align the mean of $\wclim$ with that of $\wera$. The corrected speed is thus given by
\begin{equation}
    \label{eq:mult_lin_scaling}
    \wcorr = \wraw \frac{\mu(\wera)}{\mu(\wclim)}.
\end{equation}

\subsubsection{Empirical quantile mapping}

The above methods, if applied to the historical climate projections $\wclim$, result in wind speeds with the same mean (and variance in the case of variance scaling) as the ERA-5 wind speeds. In contrast, quantile mapping achieves a precise match in the overall frequency distribution of wind speeds. This is achieved by first mapping the given wind speeds to their quantile, using the cumulative distribution function (CDF). These quantiles are than mapped back to wind speeds using the CDF of the reference dataset, in this case the ERA-5 wind speeds. The corrected wind speeds are therefore given by
\begin{equation}
    \label{eq:emp_quant_map}
    \wcorr = F^{-1}_\text{ERA-5} \left( F_\text{clim}(\wraw) \right),
\end{equation}
where $F_\text{clim}$ is the empirical CDF of the historical wind speeds from the climate dataset, and $F_\text{ERA-5}$ is the CDF of the ERA-5 wind speeds. The empirical CDFs are linearly interpolated where necessary.

\subsubsection{Weibull quantile mapping}

Weibull quantile mapping proceeds as above for empirical quantile mapping, except that the CDFs are based on Weibull distributions, which have been fitted to the historical wind speed data. The CDF of the Weibull distribution is given by
\begin{equation}
    \label{eq:weibull_cdf}
    F(w) = 1 - \text{exp}\left(-\left(\frac{w}{c}\right)^k\right),
\end{equation}
where $k$ and $c$ are the Weibull shape and scale parameters, respectively. Following \citet{li2019statistical}, we estimate these parameters using the method of moments. Their values are given by
\begin{equation}
    \label{eq:k}
    k = \left(\frac{\mu(w)}{\sigma(w)}\right)^{1.086},
\end{equation}
\begin{equation}
    \label{eq:c}
    c = \frac{\mu(w)}{\Gamma(1 + \frac{1}{k})},
\end{equation}
where $\Gamma$ is the gamma function.

\subsection{Turbine power and wake modelling}
\label{sec:wake_modelling}

Wake effects are simulated within this study using the TurbOPark engineering wake model \citep{pedersen2022turbulence}. This choice is motivated by the computational efficiency of such models, which is well suited to long-term studies, and essential to the analysis of wake effects at different length scales, which requires many simulations with different turbine configurations. The disadvantages of engineering wake models compared with alternative methods such as mesoscale numerical weather prediction (NWP) models, such as assumptions about the uniformity of the wind over the study region, are mitigated by the limited model domain extent within this study, making engineering wake models an appropriate choice. This is discussed further in section \ref{sec:discussion}.

Specifically, this paper utilises the implementation of the TurbOPark model available within the PyWake package \citep{mads_m_pedersen_2019_2562662}. The wake expansion parameter is set to 0.06. This is different to its original recommended value of 0.04 \citep{pedersen2022turbulence}, but produced better results when applied to ERA-5 data and compared with FINO1 observations (see section \ref{sec:methods_era5_validation}). This is consistent with \citet{van2023new}. We employ squared sum wake superposition. We neglect blockage effects in order to enable more computationally efficient simulation, but preliminary experiments including a blockage model showed negligible impact to the results. Since the climate models provide the wind speed at 100 \si{\metre}, we assume a power law vertical profile with an exponent of 0.11, to extrapolate to turbine hub heights. A power law profile is commonly assumed for extrapolating wind speeds from 10 \si{\metre} to hub height \citep{tobin2015assessing}, and we expect errors to be smaller here due to extrapolating from 100 \si{\metre} data. The turbines within our case study have hub heights in the range 89.5--150 \si{\metre}. The choice of an exponent of 0.11 is recommended in the literature for offshore applications \citep{hsu1994determining}.

PyWake simulations using the climate models are based on wind roses extracted at the location of the FINO1 mast. We use speed bins from 0 to 25 \si{\metre\per\second} in steps of 0.5 (plus one bin catching all speeds above 25 \si{\metre\per\second}), and direction bins of 10\si{\degree}. For a given configuration of farms, PyWake need only be run once for each combination of speed and direction. The results can then be aggregated using any arbitrary wind rose.

\begin{figure}
    \centering
    \includegraphics[width=0.7\textwidth]{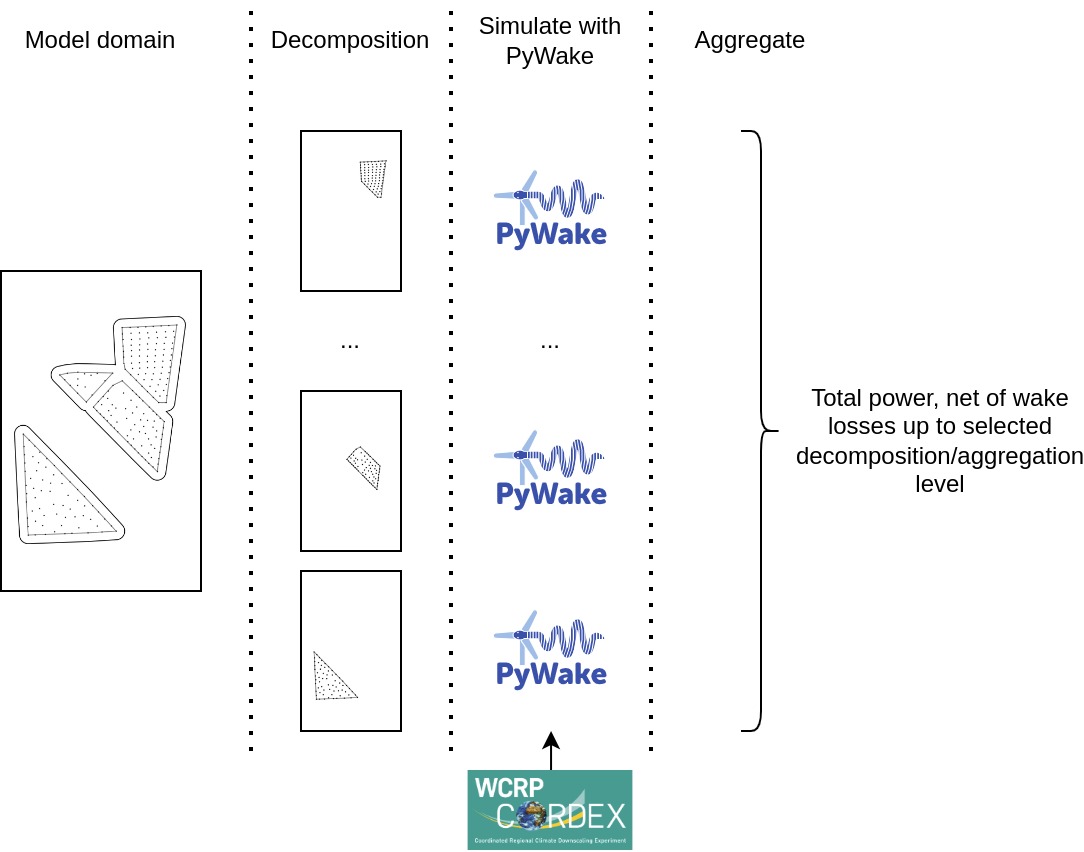}
    \caption{PyWake modelling workflow used for calculating wake losses at different spatial scales. A given simulation is based on decomposition of the model domain to a certain level. This work utilises simulations which decompose to turbine level (simulation PW1), farm level (PW2), cluster level (PW3) or no decomposition (PW4). The figure shows decomposition to the farm level. PyWake simulations use wind roses derived from downscaled climate projections from the EURO-CORDEX project, which have been bias-corrected using ERA-5 reanalysis data.}
    \label{fig:pywake_decomposition}
\end{figure}

Throughout this work, we refer to wake losses at different spatial scales. Figure \ref{fig:pywake_decomposition} outlines the modelling workflow we use to calculate these losses. To calculate total wake-induced losses for a given build-out stage (or set of farms), we perform this workflow twice. Simulation PW1 constitutes decomposition to turbine level, which is also equivalent to using no wake deficit model, or to simply extrapolating wind speeds to hub height and passing them directly through the turbine power curves. Simulation PW4 performs no decomposition, and constitutes a single PyWake simulation for the whole model domain, using the wake deficit model described above. The losses are defined as the difference in power estimated by simulations PW4 and PW1.

In some cases we attribute the total wake-induced losses to interactions at the intra-farm, intra-cluster and inter-cluster scales; see figure \ref{fig:wake_scales}. For a given build-out stage, these are estimated by running two additional sets of PyWake simulations, each following the workflow of figure \ref{fig:pywake_decomposition}. For simulation PW2, we decompose to farm level, resulting in up to 21 separate PyWake simulations (one simulation per farm), with the total power then aggregated across these simulations. The intra-farm wake-induced losses can be directly estimated from the difference between simulations PW2 and PW1. Simulation PW3 is based on decomposition to cluster level, resulting in up to eight separate PyWake simulations (one simulation per cluster; see figure \ref{fig:farms}). This enables the estimation of intra-cluster effects (the difference between simulations PW3 and PW2), and inter-cluster effects (the difference between simulations PW4 and PW3).

For the direct comparisons between ERA-5 data and FINO1 observations, the wind rose method is not sufficient, since it only allows for the calculation of high-level statistics and not a one-to-one comparison as is required for thorough validation. Instead, we take each timestamp for which we have both ERA-5 data and FINO1 observations (a total of 22,360), extract the wind speed and direction from ERA-5, and run a PyWake simulation including the farms which became operational prior to that timestamp. We assume that all turbines are continuously in operation from their installation date, and therefore neglect any downtime. For these comparisons, we run PyWake twice for each timestamp; once with no wake deficit model, and once with wakes included.

Note that where we present modelled capacity factors, either including or excluding wake effects, these are power outputs normalised by total installed capacity. However, where we quote wake losses, we define these as the difference between waked and unwaked power outputs, normalised by the unwaked power. That is, a reduction in capacity factor from 50\% to 40\% is equivalent to losses of 20\%.

\subsection{Error metrics}

We use a number of error metrics in order to make comparisons between ERA-5 and climate model data, as well as observations from the FINO1 mast. The \Rsq\, coefficient, mean bias and root mean squared error (RMSE) are all standard metrics which we use here. We additionally use the earth mover's distance (EMD) \citep{rubner1998metric}, also known as the Wasserstein metric. The EMD is a measure of the dissimilarity in the frequency distributions of two variables. Note that, since historical climate simulations are not synchronised with the historical weather, the \Rsq\, and RMSE between ERA-5 and climate datasets cannot be calculated. However, the mean bias and EMD metrics can be computed in all cases.

\subsection{Description of experiments}

\subsubsection{Validation of ERA-5 against FINO1 data}
\label{sec:methods_era5_validation}

Data from the FINO1 mast is available since April 2004, and in this work we use data up to April 2023. We divide this into two distinct time periods. The first corresponds to the period prior to the construction of the first wind farm in the region, Alpha Ventus, in April 2010. We refer to this period as the pre-AV period. We refer to the remaining time period as the post-AV period, during which time an increasing number of wind farms has become operational in the area.

The pre-AV period allows for a direct comparison between ERA-5 and FINO1 data for the purposes of validating ERA-5, without the need to account for wake effects. During the post-AV period, we expect to observe an increasing mismatch between raw ERA-5 data and FINO1 observations \citep{pettas2021effects}, unless we account for wake effects. We use PyWake to account for these effects, using the modelling approach described in section \ref{sec:wake_modelling}.

For each period, we compute the \Rsq, mean bias, RMSE and EMD error metrics between the ERA-5 data and FINO1 observations. For the post-AV period, we calculate these metrics both with and without accounting for wake effects.

\subsubsection{Benchmarking climate models against ERA-5}

Having validated both the underlying wind speeds within ERA-5, and the wake modelling approach, we can compare historical climate simulations with ERA-5. We do this over a 30-year historical period, spanning 1976--2006. 2006 marks the transition between `historical' and RCP8.5 climate simulations, and we therefore use the most recent available 30-year period prior to this point for validating the climate models.

We first investigate the wind direction distributions at the FINO1 location, according to ERA-5 and each of the seven climate models. This is important since the bias correction methods we employ within this work are applied only to the wind speed, and we therefore need to investigate the performance of the climate models in representing the overall wind direction distribution. To do this, we extract the wind rose at the FINO1 location from each climate model, and investigate the difference between the wind direction frequency distributions between each climate model, and ERA-5.

We then investigate the errors in the distribution of wind speed, wind power and simulated wake losses, as a function of climate model and bias correction method. For this, we utilise the first 20 years of the historical period (1976--1996) to train the bias correction method, and evaluate the performance on the final 10 years (1996--2006).
For wind speed distribution, we calculate the EMD and mean bias metrics. For wind power, we compute the bias in the mean and standard deviation of the simulated capacity factors, using the 2027 build-out scenario. For the wake losses, we compute the mean bias in the simulated wake-induced losses.

\subsubsection{Climate change effects}

Here we assess the changes in wind and wake effects due to projected climate change, as a function of the choice of climate model and bias correction method.

We compute the projected changes in overall wind speed and direction frequency densities, based on the downscaled climate models after bias correction with the Weibull quantile mapping method (which we find to produce the best overall results in section \ref{sec:benchmarking}). This is intended to provide insight into the consistency of the changes predicted by each climate model, as a function of the key variables influencing wake effects, \ie wind speed and wind direction.

In order to gain insight into the interaction between climate change and wake effects, we then compute the projected change in wake-induced losses between the historical and future downscaled climate datasets. For this, we again employ only the Weibull quantile mapping bias correction method, and we take only the final build-out scenario, corresponding to the year 2027. Following results in previous literature indicating that the significance of climate impacts varies by season \citep{moemken2018future,hahmann2022current}, we include such a seasonal breakdown here.

\subsubsection{Build-out effects}

Finally, we investigate how the climate and wake effects depend on farm build-out. We sort the wind farms in the region by commission date. For each unique commission date, we take all farms commissioned up to that date and run PyWake simulations based on the historical and future scenarios from each climate model. From these simulations, we then take the mean over the set of climate models, and compute the respective effects on total capacity factor of (i) climate change, (ii) intra-farm wake effects, (iii) intra-cluster wake effects, and (iv) inter-cluster wake effects. This facilitates a comparison between the magnitude of each effect, as a function of build-out.

%\FloatBarrier

\section{Results}
\label{sec:results}

\subsection{Validation of ERA-5 against FINO1 data}

\begin{figure}
    \centering
    \includegraphics[width=0.7\textwidth]{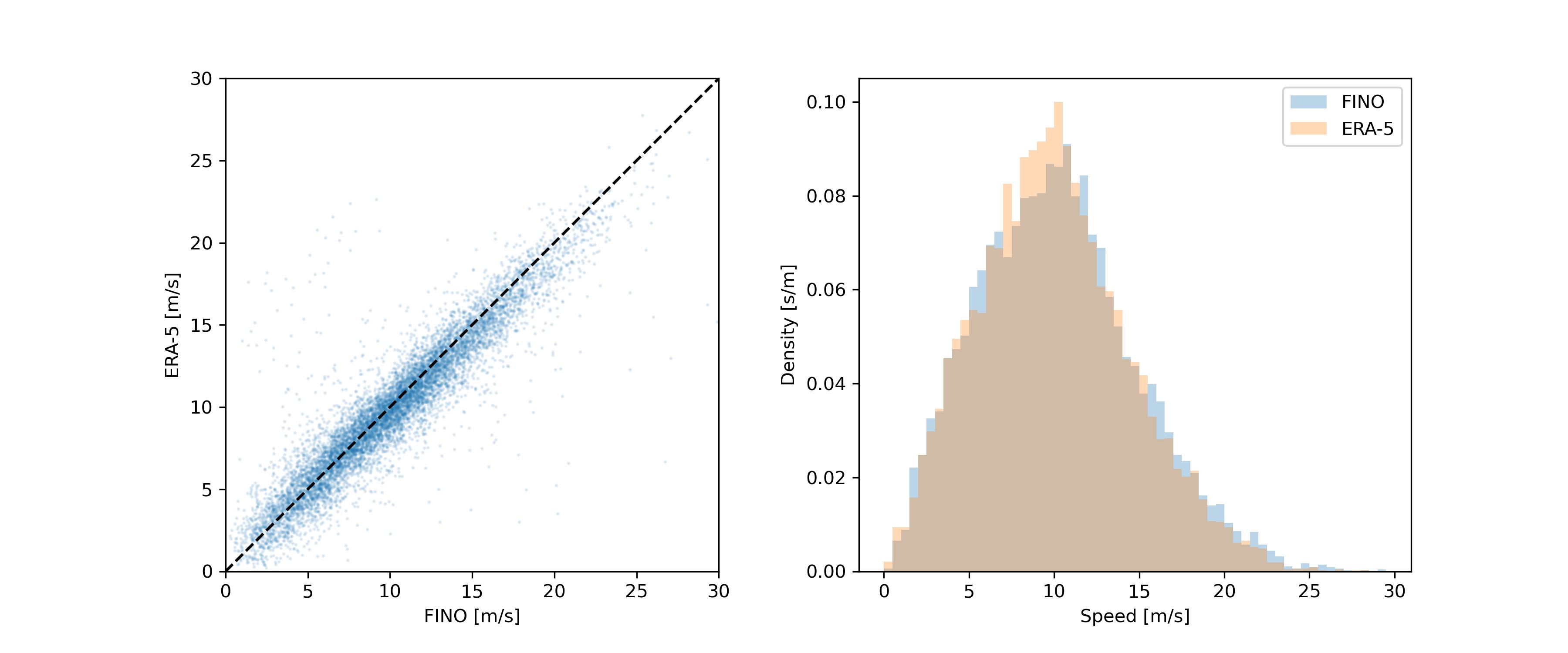}
    \caption{Comparison between ERA-5 and FINO1 100\si{\metre} wind speed, for the pre-AV period.}
    \label{fig:era_fino_pre_av}
\end{figure}

Figure \ref{fig:era_fino_pre_av} shows a comparison of 100 \si{\metre} wind speeds from ERA-5 with FINO1 observations at 102 \si{\metre}, for the pre-AV period. The corresponding error metrics are summarised in table \ref{tab:era_error_metrics}. There is a small negative bias, meaning that ERA-5 slightly underestimates wind speeds on average. The overall probability distributions agree well, as summarised by the EMD metric.

\begin{figure}
    \centering
    \includegraphics[width=0.7\textwidth]{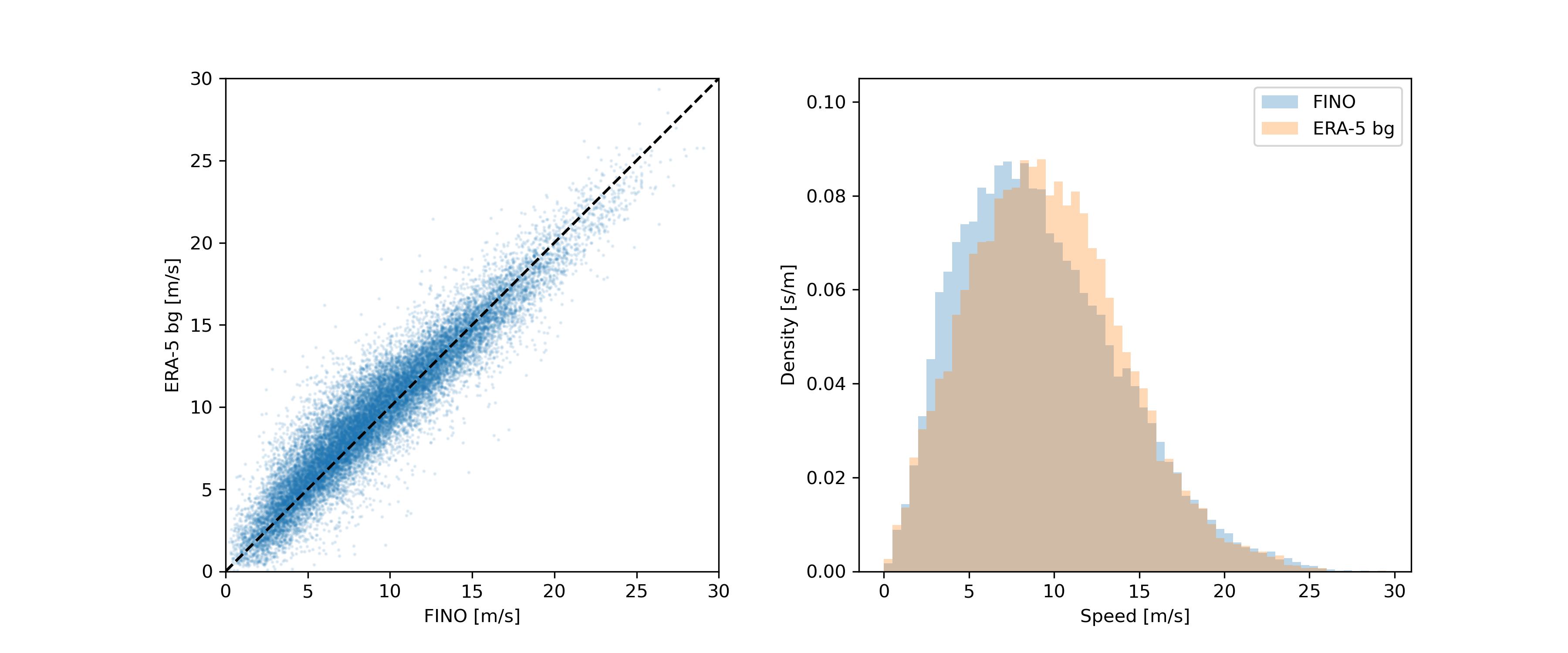}
    \includegraphics[width=0.7\textwidth]{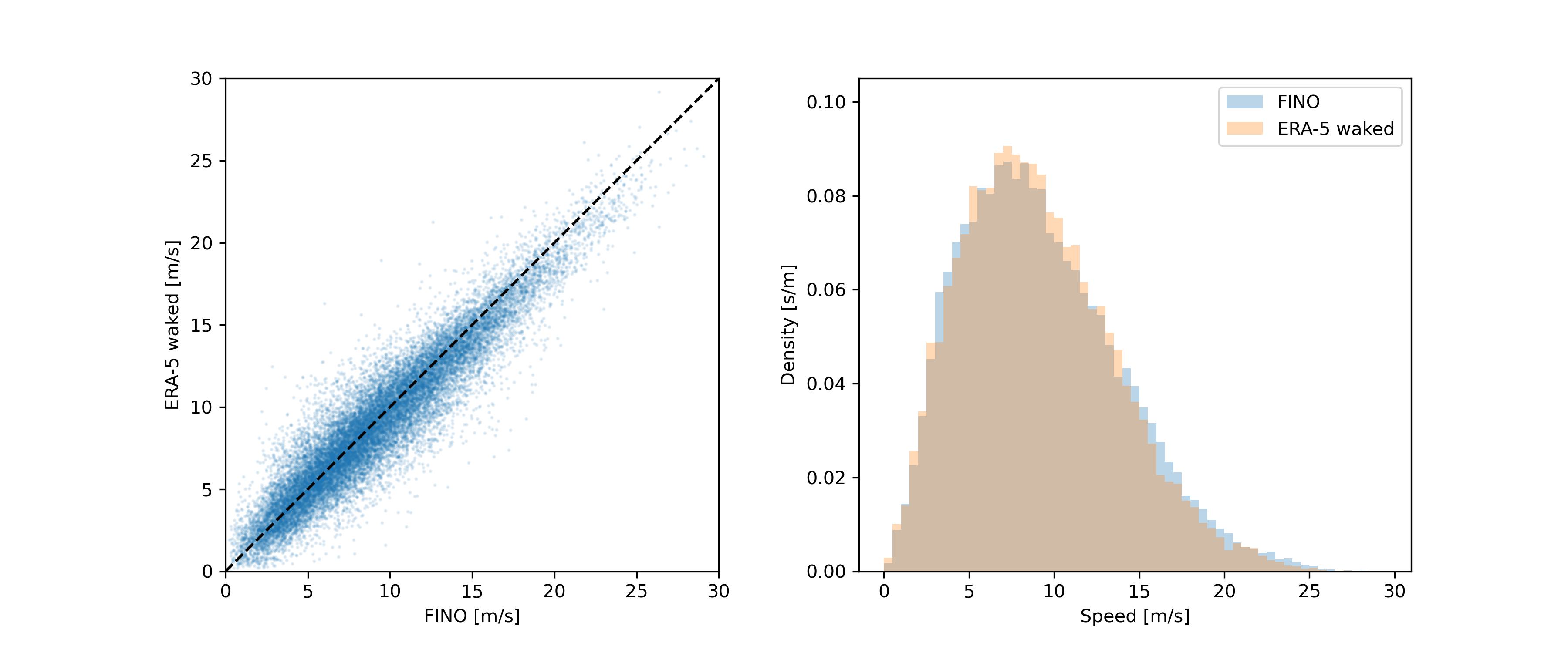}
    \caption{Comparison between ERA-5 and FINO1 100\si{\metre} wind speed, for the post-AV period. Top: ERA-5 results ignoring wake effects. Bottom: including wake effects within ERA-5.}
    \label{fig:era_fino_post_av}
\end{figure}

Equivalent results for the post-AV period are summarised in figure \ref{fig:era_fino_post_av}, and the corresponding columns of table \ref{tab:era_error_metrics}. During this period, the wind speeds measured at the FINO1 mast are increasingly influenced by wakes from the surrounding wind farms, due to its location amongst wind farms (see figure \ref{fig:farms}). Since these effects are neglected within ERA-5, we find significant discrepancies between the raw ERA-5 and FINO1 speeds. As shown in figure \ref{fig:era_fino_post_av}, ERA-5 typically overestimates wind speeds within a range of approximately 3--15 \si{\metre\per\second}, which due to the shape of typical turbine power and thrust curves corresponds to the speed range where wake effects are greatest. This is also reflected in the error metrics summarised in table \ref{tab:era_error_metrics}. While the \Rsq\, coefficient and RMSE both show good performance, there is a positive bias and a significant increase in the EMD compared with the pre-AV period.

Once we correct ERA-5 wind speeds using PyWake, we find far better agreement in terms of the overall wind speed distribution. The bias decreases, and actually takes a small negative value similar to that found for the pre-AV period, and with a significantly smaller magnitude than for the uncorrected ERA-5 speeds. The EMD decreases significantly, indicating that the overall wind speed distribution is well captured by the wake-corrected ERA-5 model. However, both the \Rsq and RMSE metrics worsen slightly compared with the uncorrected ERA-5 speeds. We attribute this to inaccuracies in the wind direction within ERA-5. The wake correction we compute depends on the wind direction, and an accurate correction therefore relies on accurate wind directions. For an individual wind field snapshot, an error in the wind direction may lead to the worsening of the agreement with observations once the wake correction is applied. This leads to an increase in the \Rsq\, and RMSE metrics, since these rely on direct comparisons on a per-timestamp basis. However, in terms of the overall wind speed distribution, errors in wind direction have the opportunity to cancel out, leading to improved overall agreement. This is reflected in the bias and EMD metric values. We also note that for applications where the overall wind resource is of primary interest, the wind speed bias and EMD metrics are likely to be of greater importance than the \Rsq\, or RMSE metrics.

\begin{table}[]
    \centering
    \begin{tabular}{c|c|c|c}
        & Pre-AV & Post-AV, ignoring wakes & Post-AV, including wakes \\ \hline
        \Rsq coefficient & 0.851 & 0.884 & 0.879 \\
        Mean bias / \si{\metre\per\second} & -0.228 & 0.390 & -0.186 \\
        RMSE / \si{\metre\per\second} & 3.31 & 2.54 & 2.64 \\
        EMD & 0.258 & 0.463 & 0.200 \\
    \end{tabular}
    \caption{Summary of error metrics between ERA-5 and FINO1 wind speeds. The error metrics are the mean bias, coefficient of determination (\Rsq), root mean squared error (RMSE) and earth mover's distance (EMD). Comparisons are presented for the periods before and after the construction of the Alpha Ventus (AV) farm. In the post-AV case, results are included for the raw ERA-5 wind speeds (ignoring wakes), and those corrected by simulating wake effects.}
    \label{tab:era_error_metrics}
\end{table}

\subsection{Benchmarking climate models against ERA-5}
\label{sec:benchmarking}

\subsubsection{Wind direction distribution}

\begin{figure}
    \centering
    \includegraphics[width=0.3\textwidth]{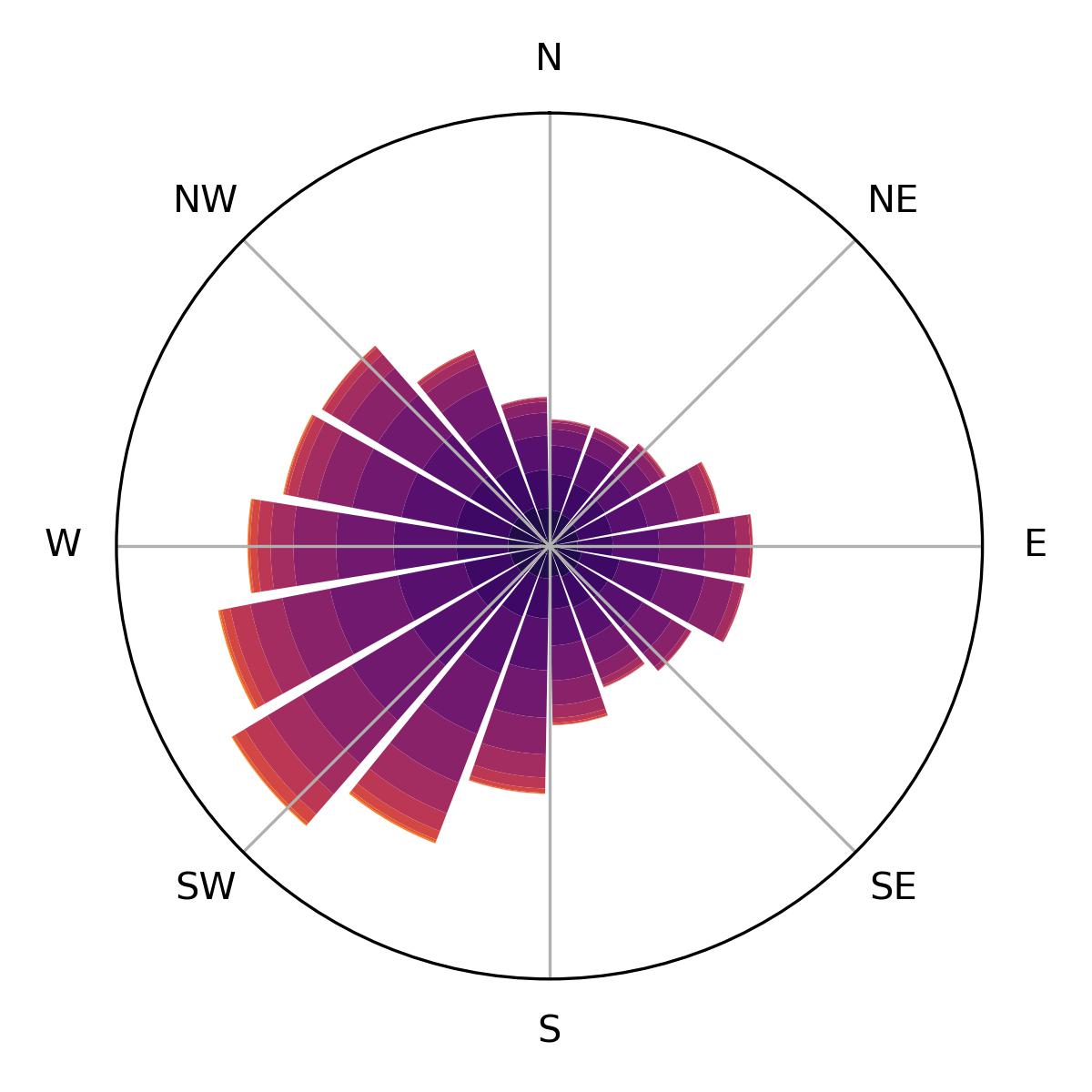}
    \caption{Wind rose at the FINO1 location, from ERA-5 data.}
    \label{fig:wind_rose_era}
\end{figure}

\begin{figure}
    \centering
    \includegraphics[width=0.8\textwidth]{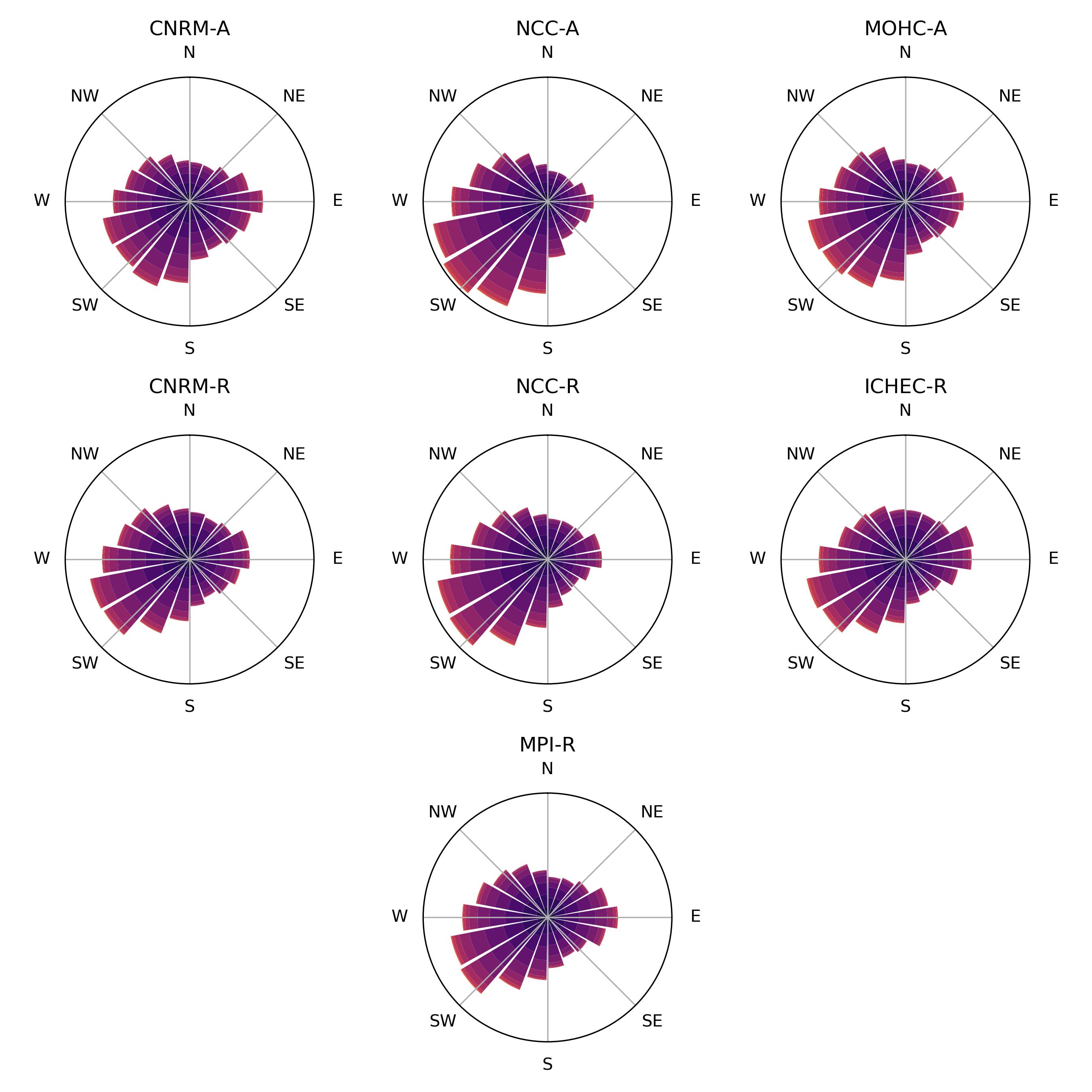}
    \caption{Wind rose at the FINO1 location, from each EURO-CORDEX model used within this study.}
    \label{fig:wind_rose_ec}
\end{figure}

\begin{figure}
    \centering
    \includegraphics[width=0.6\textwidth]{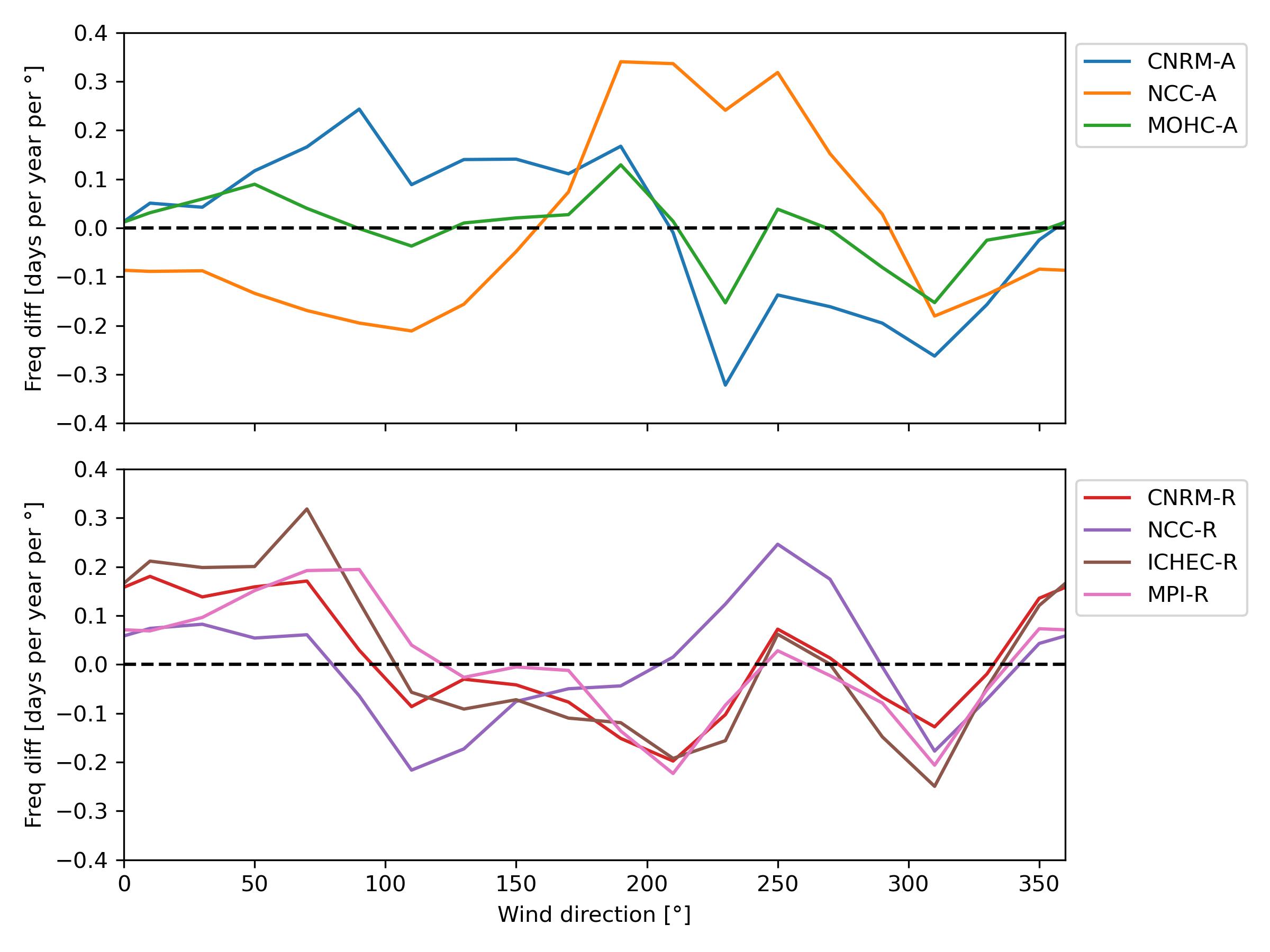}
    \caption{Difference in directional distribution between each climate model, and ERA-5, for the historical period. For legibility, the climate models are separated into groups according to the RCM used. Wind direction is measured clockwise from North, i.e. with 0 degrees corresponding to Northerly wind.}
    \label{fig:wind_rose_diffs}
\end{figure}

Figure \ref{fig:wind_rose_era} shows the wind rose from the ERA-5 dataset, at the FINO1 mast location. The prevailing wind direction is from around WSW, with very low frequency from the east sector. Figure \ref{fig:wind_rose_ec} shows the equivalent wind roses derived from each downscaled climate model. Figure \ref{fig:wind_rose_diffs} shows the differences in wind direction frequency distribution between each climate model, and the ERA-5 data. The climate model which best captures the ERA-5 wind direction distribution is MOHC-A. The worst is NCC-A, followed by CNRM-A. All four models using the RegCM4-6 RCM show a similar pattern, with reasonable agreement to ERA-5.

\subsubsection{Wind speed distribution}

Tables \ref{tab:speed_emd} and \ref{tab:speed_bias} summarise the wind speed EMD and mean bias error metric values for each combination of climate model and bias correction method, against the ERA-5 wind speeds. The CNRM-R model performs very well without any bias correction. For all other climate models, both the EMD and bias metrics can be improved by any of the bias correction methods. Variance scaling, multiplicative scaling and the two quantile mapping methods share very similar performance for both metrics, and perform well. The offset bias correction performs worse but is still able to improve the EMD metric for all climate models. The MPI-R climate model is improved only slightly by any of the bias correction methods.

\begin{table}[]
    \centering
    \small
    \begin{tabular}{C{3cm}|C{1.2cm}|C{1.2cm}|C{1.2cm}|C{1.2cm}|C{1.2cm}|C{1.2cm}}
        \diagbox[height=5cm,width=3.4cm]{\raisebox{3ex}{Climate model}}{\rotatebox{90}{Bias correction method}} & \rotatebox{90}{None} & \rotatebox{90}{Offset} & \rotatebox{90}{Variance scaling} & \rotatebox{90}{Multiplicative scaling} & \rotatebox{90}{Empirical quantile mapping} & \rotatebox{90}{Weibull quantile mapping} \\ \hline
        CNRM-A & 0.33 & 0.18 & 0.11 & 0.12 & \textbf{0.1} & \textbf{0.1} \\
        NCC-A & 0.68 & 0.17 & \textbf{0.09} & \textbf{0.09} & 0.1 & \textbf{0.09} \\
        MOHC-A & 0.45 & 0.27 & 0.25 & \textbf{0.18} & \textbf{0.18} & \textbf{0.18} \\
        CNRM-R & \textbf{0.07} & 0.08 & 0.09 & \textbf{0.07} & 0.08 & 0.08 \\
        NCC-R & 0.3 & 0.11 & 0.11 & \textbf{0.08} & 0.1 & 0.1 \\
        ICHEC-R & 0.37 & 0.08 & 0.09 & 0.09 & \textbf{0.07} & 0.09 \\
        MPI-R & 0.39 & 0.3 & \textbf{0.29} & 0.3 & 0.32 & 0.31 \\
    \end{tabular}
    \caption{Wind speed EMD values for each combination of climate model and bias correction method. Units: \si{\metre\per\second}. Bold values indicate the lowest value in each row.}
    \label{tab:speed_emd}
\end{table}

\begin{table}[]
    \centering
    \small
    \begin{tabular}{C{3cm}|C{1.2cm}|C{1.2cm}|C{1.2cm}|C{1.2cm}|C{1.2cm}|C{1.2cm}}
        \diagbox[height=5cm,width=3.4cm]{\raisebox{3ex}{Climate model}}{\rotatebox{90}{Bias correction method}} & \rotatebox{90}{None} & \rotatebox{90}{Offset} & \rotatebox{90}{Variance scaling} & \rotatebox{90}{Multiplicative scaling} & \rotatebox{90}{Empirical quantile mapping} & \rotatebox{90}{Weibull quantile mapping} \\ \hline
        CNRM-A & 0.32 & 0.05 & \textbf{0.04} & 0.06 & 0.06 & 0.06 \\
        NCC-A & 0.68 & 0.06 & \textbf{0.04} & 0.07 & 0.08 & 0.07 \\
        MOHC-A & 0.45 & -0.2 & -0.25 & -0.18 & \textbf{-0.17} & -0.18 \\
        CNRM-R & \textbf{0.04} & 0.07 & 0.08 & 0.06 & 0.07 & 0.07 \\
        NCC-R & 0.3 & \textbf{0.05} & \textbf{0.05} & 0.06 & 0.06 & 0.06 \\
        ICHEC-R & 0.37 & -0.03 & -0.05 & \textbf{-0.02} & \textbf{-0.02} & \textbf{-0.02} \\
        MPI-R & 0.39 & 0.3 & \textbf{0.29} & 0.3 & 0.32 & 0.31 \\
    \end{tabular}
    \caption{Mean wind speed bias values for each combination of climate model and bias correction method. Units: \si{\metre\per\second}. Bold values indicate the lowest magnitude value in each row.}
    \label{tab:speed_bias}
\end{table}

\subsubsection{Wind power and wake loss distributions}

Table \ref{tab:bg_pow_mean_bias} summarises the biases in the mean simulated capacity factor between the (bias-corrected) climate datasets and ERA-5, ignoring all wake effects. Table \ref{tab:bg_pow_std_dev_bias} shows the equivalent bias in the simulated capacity factor standard deviation. For reference, the simulated capacity factor value for the ERA-5 dataset is 58\%, with a standard deviation of 38\%. Note that this is the `unwaked' capacity factor, hence its value is larger than typical observed capacity factors.

The CNRM-R model produces very accurate estimates of capacity factor and its standard deviation; this is consistent with its performance for the wind speed distribution. All other climate datasets overestimate the mean capacity factor by varying amounts up to around 4\%, but can be improved by any of the bias correction methods. The MPI-R model improves only very slightly with bias correction; this is again consistent with the results from analysing the wind speed distribution.

Regarding the standard deviation of the capacity factor, the results do not favour a particular bias correction method or climate model. No single bias correction model is able to improve the biases of all climate models. However, all of the biases are small compared with the ERA-5 standard deviation of 38\%, indicating that all models capture the variability in wind power to a reasonable degree.

Finally, table \ref{tab:loss_mean_bias} summarises the biases in the mean simulated wake-induced reduction in capacity factor. The CNRM-R and ICHEC-R models show very small bias, but all of the other climate models can be improved by bias correction. However, there is no clear advantage to any particular model or bias correction method, and no single bias correction method is able to improve all seven climate models. We again attribute this to the fact that all of these biases are small compared with the ERA-5 wake losses, which are 20.2\%.

Considering the evaluation metrics for wind speed, capacity factor and wake losses, we suggest that Weibull quantile mapping performs best overall. This is consistent with findings from previous literature \citep{li2019statistical}. For the remaining results, we therefore focus on projections corrected using Weibull quantile mapping. Note however that we retrain the bias correction on the entire historical period (1976--2006).

\begin{table}[]
    \centering
    \small
    \begin{tabular}{C{3cm}|C{1.2cm}|C{1.2cm}|C{1.2cm}|C{1.2cm}|C{1.2cm}|C{1.2cm}}
        \diagbox[height=5cm,width=3.4cm]{\raisebox{3ex}{Climate model}}{\rotatebox{90}{Bias correction method}} & \rotatebox{90}{None} & \rotatebox{90}{Offset} & \rotatebox{90}{Variance scaling} & \rotatebox{90}{Multiplicative scaling} & \rotatebox{90}{Empirical quantile mapping} & \rotatebox{90}{Weibull quantile mapping} \\ \hline
        CNRM-A & 2.67 & 0.6 & \textbf{0.16} & 0.68 & 0.28 & 0.3 \\
        NCC-A & 3.99 & -0.07 & -0.17 & 0.63 & -0.19 & \textbf{-0.06} \\
        MOHC-A & 2.28 & -2.12 & -2.23 & -1.53 & -1.54 & \textbf{-1.52} \\
        CNRM-R & \textbf{-0.03} & 0.41 & 0.34 & 0.39 & 0.36 & 0.2 \\
        NCC-R & 1.3 & -0.27 & -0.21 & \textbf{0.13} & -0.19 & -0.28 \\
        ICHEC-R & 2.34 & -0.55 & -0.53 & \textbf{-0.22} & -0.29 & -0.31 \\
        MPI-R & 2.73 & 1.95 & \textbf{1.8} & 1.9 & 1.98 & 1.98 \\
    \end{tabular}
    \caption{Capacity factor mean bias values for each combination of climate model and bias correction method. Units: \%. Bold values indicate the lowest magnitude value in each row.}
    \label{tab:bg_pow_mean_bias}
\end{table}

\begin{table}[]
    \centering
    \small
    \begin{tabular}{C{3cm}|C{1.2cm}|C{1.2cm}|C{1.2cm}|C{1.2cm}|C{1.2cm}|C{1.2cm}}
        \diagbox[height=5cm,width=3.4cm]{\raisebox{3ex}{Climate model}}{\rotatebox{90}{Bias correction method}} & \rotatebox{90}{None} & \rotatebox{90}{Offset} & \rotatebox{90}{Variance scaling} & \rotatebox{90}{Multiplicative scaling} & \rotatebox{90}{Empirical quantile mapping} & \rotatebox{90}{Weibull quantile mapping} \\ \hline
        CNRM-A & \textbf{0.01} & 0.41 & 0.16 & 0.06 & 0.08 & 0.07 \\
        NCC-A & -0.32 & 0.33 & \textbf{-0.01} & -0.28 & \textbf{-0.01} & -0.16 \\
        MOHC-A & 0.07 & 0.66 & -0.16 & \textbf{-0.01} & -0.3 & -0.25 \\
        CNRM-R & -0.3 & -0.46 & -0.29 & -0.39 & \textbf{-0.17} & -0.31 \\
        NCC-R & -0.26 & -0.12 & -0.1 & -0.34 & \textbf{-0.05} & -0.15 \\
        ICHEC-R & -0.36 & \textbf{0.12} & -0.22 & -0.31 & -0.19 & -0.29 \\
        MPI-R & -0.82 & -0.59 & -0.76 & -0.73 & \textbf{-0.55} & -0.8 \\
    \end{tabular}
    \caption{Capacity factor standard deviation bias values for each combination of climate model and bias correction method. Units: \%. Bold values indicate the lowest magnitude value in each row.}
    \label{tab:bg_pow_std_dev_bias}
\end{table}

\begin{table}[]
    \centering
    \small
    \begin{tabular}{C{3cm}|C{1.2cm}|C{1.2cm}|C{1.2cm}|C{1.2cm}|C{1.2cm}|C{1.2cm}}
        \diagbox[height=5cm,width=3.4cm]{\raisebox{3ex}{Climate model}}{\rotatebox{90}{Bias correction method}} & \rotatebox{90}{None} & \rotatebox{90}{Offset} & \rotatebox{90}{Variance scaling} & \rotatebox{90}{Multiplicative scaling} & \rotatebox{90}{Empirical quantile mapping} & \rotatebox{90}{Weibull quantile mapping} \\ \hline
        CNRM-A & -1.67 & -2.91 & -1.74 & -0.98 & \textbf{-0.86} & -1.06 \\
        NCC-A & -4.05 & -3.68 & -1.75 & \textbf{0.38} & -1.83 & -0.88 \\
        MOHC-A & -4.59 & -5.2 & \textbf{-0.07} & -0.58 & 1.47 & 0.85 \\
        CNRM-R & \textbf{-0.02} & 1.37 & 0.28 & 1.06 & 0.05 & 0.36 \\
        NCC-R & -1.94 & -1.37 & -1.38 & \textbf{0.26} & -1.05 & -1.14 \\
        ICHEC-R & \textbf{-0.01} & -0.68 & 1.55 & 2.04 & 1.36 & 1.88 \\
        MPI-R & 1.91 & 1.14 & 2.0 & 1.91 & \textbf{0.26} & 2.23 \\
    \end{tabular}
    \caption{Wake loss mean bias values for each combination of climate model and bias correction method. Units: \%. Bold values indicate the lowest magnitude value in each row.}
    \label{tab:loss_mean_bias}
\end{table}

\subsection{Climate change effects}

\subsubsection{Change in wind distribution}

\begin{figure}
    \centering
    \includegraphics[width=0.8\textwidth]{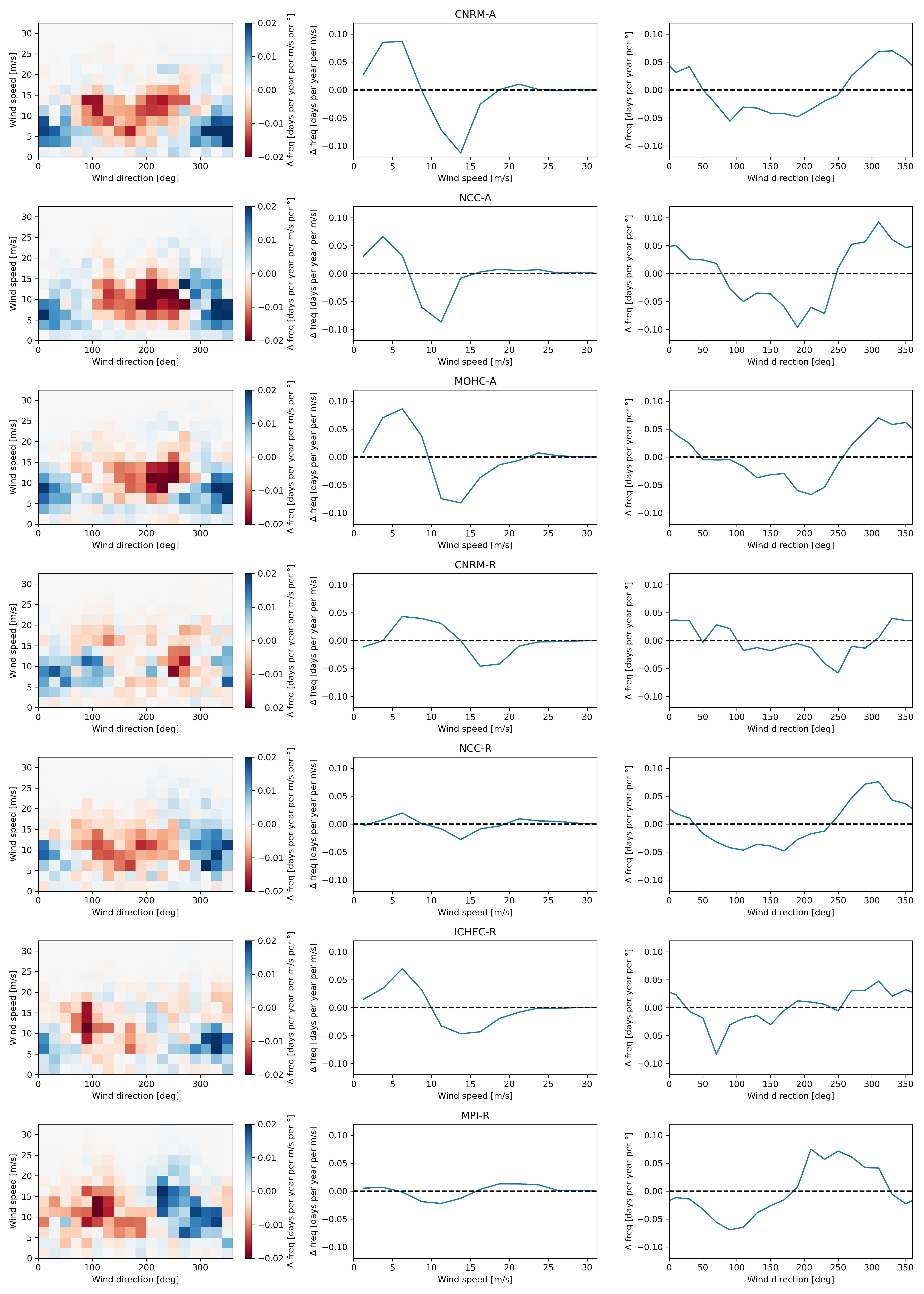}
    \caption{Projected change in wind speed and direction distributions, from each climate model, after bias correction using Weibull quantile mapping.}
    \label{fig:wind_change_with_climate}
\end{figure}

Figure \ref{fig:wind_change_with_climate} shows the projected change in wind speed and direction distribution at the FINO1 mast, between the historical and future periods, according to each bias-corrected climate model (using the Weibull quantile mapping method).

Common to all models is an increase in the projected frequency of low wind speeds, a decrease in frequency of medium speeds, and very little change to the frequency of high speeds. However, the transition point between positive and negative differences varies between models. For the CNRM-R model, the transition point is the highest, at almost 15 \si{\metre\per\second}. For the CNRM-A, MOHC-A and ICHEC-R models, the transition is at around 10 \si{\metre\per\second}, while for the NCC-A model, the transition is lower, at around 7.5 \si{\metre\per\second}. The NCC-R and MPI-R models show only a very weak signal.

All models indicate an increase in frequency for approximately half of the wind directions, and a corresponding decrease in the other half. However, the magnitude of the changes, as well as the dividing lines between increasing and decreasing frequency, again vary between each model. Most models (with MPI-R the exception) show an increase in frequency of northerly winds, but there is little agreement in the change in easterly, southerly or westerly winds.

These results may have implications for wake mitigation efforts. The lack of fine-scale agreement on wind speed and direction distributions may prevent the design of specific farm layouts which are robust to future changes in wind climate, although it is more likely that larger-scale design choices such as turbine sizes and densities can be informed by such projections.

\subsubsection{Change in power output and wake effects}
\label{sec:results_climate_power}

\begin{figure}
    \centering
    \includegraphics[width=0.8\textwidth]{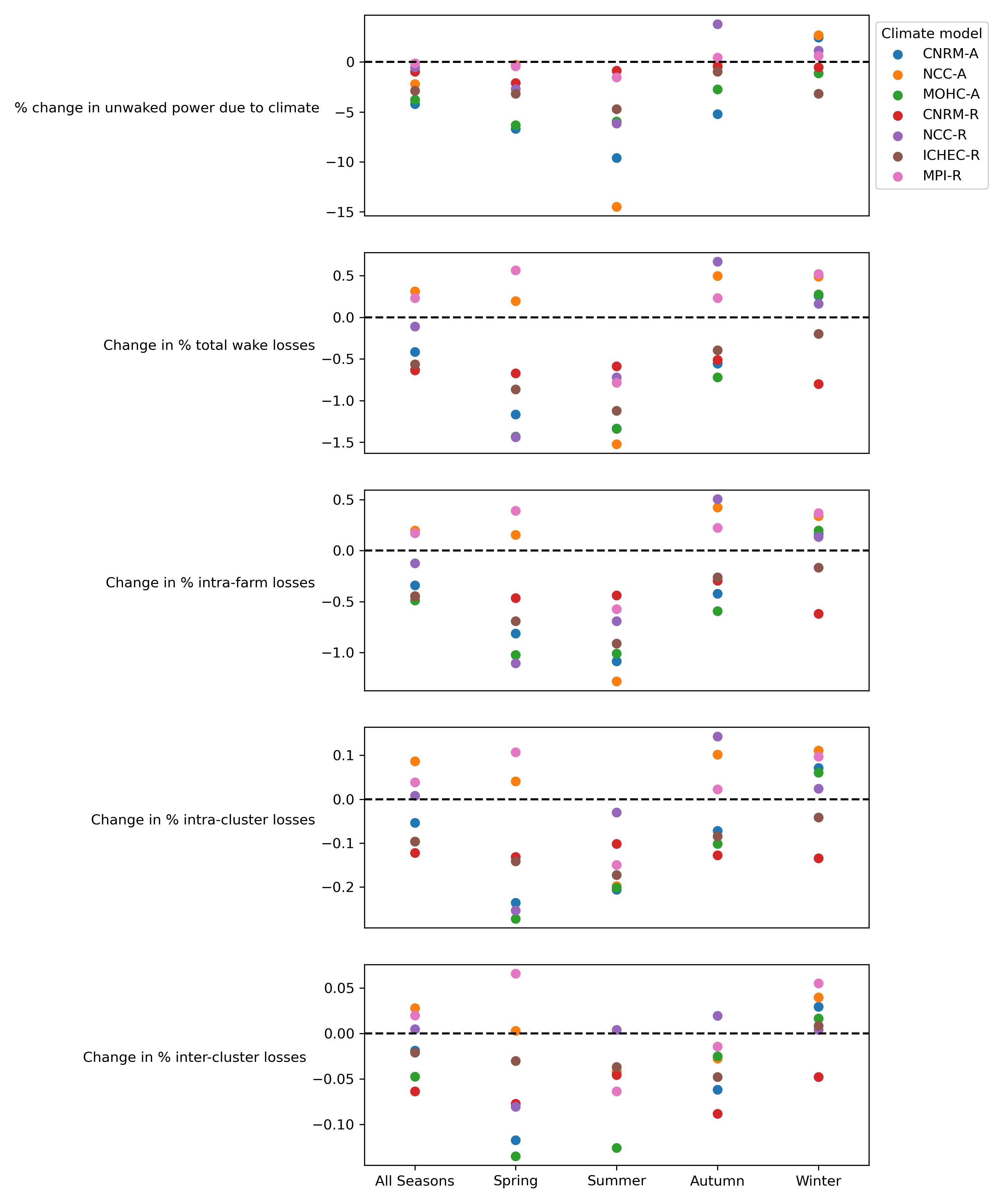}
    \caption{Projected changes in unwaked power, intra-farm losses, intra-cluster losses, inter-cluster and total losses, due to climate change. The change in unwaked power is expressed as a percentage of the `historical' unwaked power. Changes in wake losses are absolute changes in the \% losses as calculated for the historical and future periods. Each effect is broken down by season. All results utilise the Weibull quantile mapping bias correction.}
    \label{fig:changes_with_climate}
\end{figure}

Figure \ref{fig:changes_with_climate} shows the change in predicted unwaked power and wake-induced losses, due to climate change. We include results from each downscaled climate model, using Weibull quantile mapping bias correction, and include both full results and a breakdown by season.

The unwaked power is more sensitive than the wake effects to climate change. For the all-season results, all models predict a decrease in mean power, of up to 4\%. The spread of climate projections for individual seasons is much greater. All models predict decreasing power for spring and summer, but there is no consensus for autumn or winter. The increased significance of changes in summer is consistent with previous literature \citep{carvalho2017potential,moemken2018future}. The overall predicted reduction in wind resource has potential consequences for (robust) future wind farm designs. For example, cost-effective wind farm design may require larger and lower density farms in order to compensate for reduced resource density.

There is no overall consensus on the sign of the all-season total wake-induced losses. Four of the models show negative change in wake-induced losses (that is, an increase in the magnitude of the losses). These models are CNRM-A, MOHC-A, CNRM-R and ICHEC-R. This is consistent with the results of figure \ref{fig:wind_change_with_climate}; these four models exhibit increased frequency of all wind speeds below 10 \si{\metre\per\second}. Wind turbine power curves are most sensitive to the wind speed within the cubic portion of the power curve, \ie above the cut-in speed but below the rated speed. This roughly corresponds to around 3 to 10 \si{\metre\per\second} for typical turbines, resulting in increased wake-induced losses for these models. The two models with the weakest signal in figure \ref{fig:wind_change_with_climate} are NCC-R and MPI-R, and we correspondingly find that these two models predict the lowest magnitude changes in mean wake-induced losses. For the NCC-A model, the projected increase in frequency of wind speeds below 7.5 \si{\metre\per\second} would be expected to increase the wake-induced losses at very low speed, but the decrease in frequency between 7.5 and 15 \si{\metre\per\second} outweighs this effect, resulting in smaller wake-induced losses overall (\ie a positive change). These results highlight the sensitivity of projected changes in wake-induced losses to the precise distribution of wind speeds, and the challenges in projecting such effects in future climates.

During the summer, all climate models predict an increase in the magnitude of total wake-induced losses. Looking at the break-down into wakes at different scales, both intra-farm and intra-cluster losses are projected to increase during summer, although the change in inter-cluster wakes is not consistent across all climate models. The overall increasing magnitude of wake-induced losses during summer compounds the effects of seasonal variation and projected climate change.

The spread in the predicted changes across the climate models highlights the need for model ensembles, in order to capture the high uncertainty inherent in projecting climate change. This is particularly the case for the direct climate effect, which is of greater magnitude than changes in wake effects, and has a spread of over 4\%. In contrast, the spread in projected changes in total wake-induced losses is smaller, but is still more than 1\%.

\subsection{Build-out effects}
\label{sec:results_build_out_effects}

\begin{figure}
    \centering
    \includegraphics[width=0.8\textwidth]{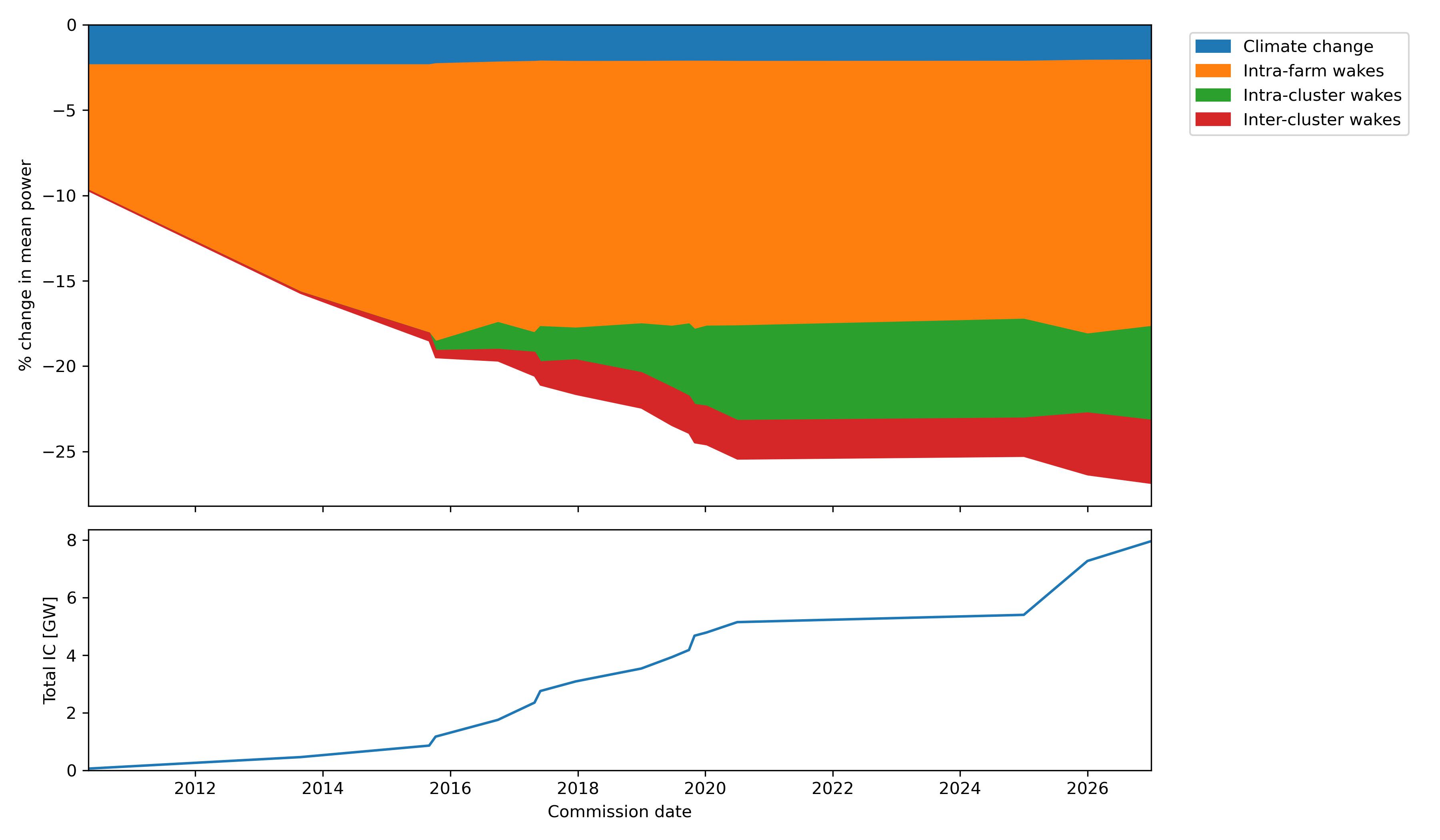}
    \caption{Top: evolution of climate and wake effects as a function of wind farm build-out date, expressed as percentage changes in the mean power output. Bottom: cumulative installed capacity (IC). Note that the dates on the horizontal axis correspond to farm commission dates, indicating the farms included in the model. The simulation periods are held fixed.}
    \label{fig:climate_wake_effects}
\end{figure}

A comparison of climate change and wake effects, as a function of wind farm build-out in the region, is shown in figure \ref{fig:climate_wake_effects}. Note that here we have taken the average effects over all seven climate models, which were bias-corrected using the Weibull quantile mapping method. We discuss each effect in turn.

The climate effect depends only very weakly on the build-out stage, holding a value of approximately 2.1\%. This is because we are assuming constant wind speeds over the region for the purposes of the PyWake simulations, and the simulation periods are the same for all build-out stages. Any change in climate effect with build-out therefore arises only due to the use of different power curves for each new wind farm, resulting in a negligible effect.

Intra-farm effects initially grow with build-out, before stabilising. The first commissioned farm, Alpha Ventus, is very small, consisting of only 12 turbines, and therefore incurs relatively low intra-farm wake-induced losses. Most of the subsequent farms are of similar size and turbine spacing, and therefore the average intra-farm losses stabilise. The intra-farm losses at the final build-out stage are 13.3\%.

Intra-cluster wake effects are zero by definition when only one farm is present, but then grow with increasing build-out. This is due to the introduction of additional wind farms within several clusters. Cluster B, for example, consists of five separate farms, each of which became operational at a different time, leading to increasing intra-cluster effects. The intra-cluster losses are 4.3\% by the final build-out stage.

Finally, inter-cluster wake effects also grow with build-out. They are again zero by definition when there is only one cluster present. However, as the number if clusters grows, and they become larger and more closely spaced, inter-cluster effects increase, reaching a final value of 2.5\%.

By the 2027 build-out, both intra-cluster and inter-cluster wake-induced losses are therefore of greater magnitude than the mean long-term climate change effect, although as shown in figure \ref{fig:changes_with_climate} it should be noted that some climate models predict changes of up to 4\%. However, it is also important to note that the predicted climate change effect is based on climates approximately 100 years apart, and assuming the RCP8.5 scenario, which is a high-emission scenario. In contrast, the build-out evolution we have assessed spans less than two decades, and could be considered a modest scenario compared with the potential longer-term build-out, including several farms in the study region which are already in the planning pipeline \citep{wind_farm_data}. This leads to the conclusion that increased losses due to intra-cluster and inter-cluster wake effects will outgrow changes due to the climate.

\section{Discussion}
\label{sec:discussion}

The results of this study indicate that both climate change and wake effects constitute significant sources of long-term change in wind farm power output. Owing to farm lifetimes of 25--30 years, and offshore lease durations of 60 years, both of these factors should be considered at the wind farm planning stage.

The results of section \ref{sec:results_climate_power} indicate that the magnitude of wake effects differs between the historical and future climates, implying an interaction between climate and wake effects. Looking across all seasons, all climate models agree that there is a reduction in mean wind speeds at this location, or equivalently an increase in frequency of lower speeds and a decrease in frequency of higher speeds. Wake effects are strongest within a fairly narrow wind speed range, which is dictated by the cut-in and rated speeds of the turbines. This means that changes in wake effects are determined by changes in the occurrence frequency of specific wind speeds. This leads to high sensitivity, and a lack of consensus between each climate model on the overall change in wake effects under the future climate scenario. However, isolating the summer months, all climate models agree that there will be an increase in the magnitude of wake-induced losses. This compounds the effects of seasonal variation (with lowest winds in summer), as well as the direct climate effect of weakening wind resources in the region. Given that wind farms are long-term investments, these results suggest that it is important to consider both climate change and wake effects at all spatial scales within wind farm planning.

The results of this study may inform the development of robust approaches to long-term farm planning and design. We note that there is limited agreement among the climate models in terms of which precise wind speeds and directions change in frequency in future. This lack of fine-scale agreement may prevent the design of specific farm layouts which are robust to future changes in wind climate, although it is likely that larger-scale design choices such as turbine sizes and densities can be informed by such projections, with lower densities and larger turbines likely to be favoured by reduced resource density in future. For questions of future wind farm planning and design, distinguishing between intra-farm, intra-cluster and inter-cluster wake effects is particularly useful. Intra-farm losses can always be mitigated via array layout optimisation when designing an individual farm \citep{PIGGOTT2022176}, and optimal control once the farm is operational. There may be some scope to reduce intra-cluster losses via similar mechanisms, particularly if multiple farms within the same cluster are being designed simultaneously and by the same developer, or will be operated cooperatively. From the perspective of a wind farm developer, inter-cluster losses will be the most challenging to mitigate, but from the perspective of policy-makers, these losses may be the most important for informing lease area valuation and long-term planning decisions.

We chose to use an engineering wake modelling approach within this work. The wind inputs to these models were wind roses derived from downscaled climate projections from the EURO-CORDEX project, with a resolution of 0.11$^\circ$, or around 12.5 \si{\kilo\metre}. There are several advantages to this engineering wake modelling approach, which enable the analysis of key drivers of long-term change in wind power outputs. It is computationally efficient, which is essential to running the large number of simulations required to isolate wakes effects at different spatial scales, and the use of wind roses makes it straightforward to calculate average power and wake effects under different climate scenarios. A disadvantage of this method is the introduction of unknown hyperparameters, in this case the wake expansion parameter. Previous work has shown that the optimal value of this parameter may vary with wind conditions, and the spatial scale of the wakes being modelled \citep{van2024scalable}. However, the comparisons we presented with observations from the FINO1 mast indicate that our model setup performs adequately. Alternative modelling approaches would include mesoscale NWP modelling, or CFD methods such as RANS or LES. Mesoscale modelling in particular is often used for long-term wind power assessments, and would have two main advantages over the engineering modelling approach. Firstly, the use of mesoscale models would further downscale the climate projection datasets, potentially producing a more accurate representation of the local wind conditions over the study site, and incorporating a wider variety of atmospheric variables into the estimation of wind farm power. Secondly, it would remove the assumption that wind fields are uniform over the model domain. Mesoscale spatial variations may be significant in study regions such as the German Bight \citep{von2022investigation}. However, the computational cost of mesoscale NWP modelling would be prohibitive for the analysis performed within this work, where the focus is on long-term behaviour, and the wake analysis requires a large number of individual simulations with different turbine configurations. The use of engineering wake modelling within this study is therefore a practical choice, which is motivated by the extensive validation of the TurbOPark model in the literature \citep[\eg][]{stieren2021evaluating,pedersen2022turbulence,nygaard2022large,van2024performance,sorensen2024extension,zum2024evaluation}, as well as the validation against FINO1 observations we performed within this study.

Assessing future changes due to either climate or wake effects carries significant uncertainty. Within this work, an ensemble of seven downscaled climate projections produces a range of responses in wind power for the study region, from around 0 to 4\%. This is despite all of the climate projections relying on the same RCP8.5 emissions pathway. As described in \ref{sec:alt_climates}, simulations based on an alternative lower-emission scenario (RCP2.6) produce climate-induced changes in mean power which differ in magnitude compared with RCP8.5, although with a lack of agreement between the climate models as to which climate scenario produces the largest changes. Simulations for an intermediate future period (2040--2070) show a slight reduction in the mean climate impact compared with the 2070--2100 period, although again with a lack of agreement between different climate models. Overall, the results of \ref{sec:alt_climates} demonstrate the high uncertainty inherent in using climate projections to model future wind power production. With respect to wake effects, this study is based on known or planned turbine locations and capacities. Nevertheless, as demonstrated by the experiments summarised in \ref{sec:sensitivity_to_ct}, uncertainty in precise turbine characteristics introduces a significant amount of uncertainty in the simulated wake losses. While this study has therefore demonstrated that changes in both climate and wake effects are impactful, further work is required to investigate the relative sizes of the uncertainties associated with each.

Uncertainty in wake effects will increase for build-out time horizons beyond those considered in this study. For example, future work could incorporate much larger wind farms likely to be developed in the coming decades (\eg based on future scenarios such as \citet{waldman2023wind}). However, the details of such scenarios, including their associated turbine design parameters, are highly speculative, and their simulation will require an appropriate analysis and treatment of the resulting uncertainties. Furthermore, as discussed above, expanding the study area may stretch the validity of the current modelling approach (which assumes spatially and temporally uniform wind fields), and instead require mesoscale modelling to correctly capture spatial variation across the domain. An intermediate option could be to use streamline-following engineering wake models \citep[e.g.][]{von2022investigation}. It is for these reasons that we chose to limit the present study to farms with known designs. Although this results in a discrepancy between the build-out and climate change timescales investigated, we are still able to draw conclusions about the relative impacts of these effects. Further exploration of the impact of future build-out at larger scales is left to future work.

For the study region of this paper, the change in wake-induced losses from two decades of build-out is of significantly greater magnitude than the change in power due to approximately 100 years of climate change under the RCP8.5 scenario. Therefore, while the impact of climate change on future wind power is well studied in the literature, the results of this study highlight the importance of long-term build-out as a key driver of change in wind power production. However, we also note that the German Bight is a region with a particularly high density of wind farms, while the impact of climate change on wind resources in the North Sea is relatively small compared with other regions worldwide \citep{warder2024mapping}. There would be value in future work repeating the analysis of this study for other regions, where the relative magnitudes of the climate and wake effects will likely differ. However, this would rely on the availability of data specifying turbine locations, characteristics (power, rotor area, etc) and installation dates, which may be a limiting factor in extending this analysis globally.

\section{Conclusions}
\label{sec:conclusions}

In this study we have investigated and compared the potential effects of climate change and increasing wake-induced losses due to farm build-out, for a study region containing 21 wind farms in the German Bight. Our analysis was based on an ensemble of seven EURO-CORDEX climate projections, which we bias-corrected using ERA-5 reanalysis data. Wake effects were accounted for using the TurbOPark wake deficit model available within PyWake, which we validated via comparisons with observations from the FINO1 mast, before and after farms became operational in the area.

Regarding the bias correction, we found that variance scaling, multiplicative scaling, empirical quantile mapping and Weibull quantile mapping methods all produced good results. All seven climate models were able to simulate wind farm capacity factors to within 2\% compared with ERA-5, after bias correction with Weibull quantile mapping. Both capacity factor variability and mean wake losses were well captured even without the application of bias correction.

Applying the Weibull quantile mapping bias correction method to downscaled future climate projections for the period 2070--2100, we then investigated and compared the impacts of climate change and wake effects, the latter evolving as a function of wind farm build-out in the region between 2010 and 2027. Looking firstly at the climate effect, all climate models project decreases in wind power for the final 21-farm scenario for the 2070--2100 period compared with 1976--2006, with the magnitude of the change varying between 0.03 and 4.2\%. Looking instead at wake effects and averaging over all climate models for the 2070--2100 period, we found that intra-cluster and inter-cluster wake effects grow from zero (by definition) when only one wind farm is present, to 4.3 and 2.5\% respectively for the full 21-farm build-out scenario, representing around two decades of build-out. Intra-farm wake-induced losses are less sensitive to the build-out, but are larger in magnitude, at 13.2\% for the final build-out. Wake effects between farms and clusters have typically been underestimated in the context of resource assessment, but our results indicate that their impact after less than two decades of build-out is of greater magnitude than a century of climate change under the high-emissions RCP8.5 scenario. These wake effects are likely to continue to grow with further build-out, and it is vital to include their effects in long-term wind farm planning.

Our results indicate that wake-induced losses differ between the historical and future periods, under the RCP8.5 climate scenario. This implies an interaction between the climate and wake impacts. In spring and summer, wake-induced losses are found to be of greater magnitude under the future scenario, with the greatest change occurring during summer. This effect compounds existing seasonal variability in wind power, and exacerbates the negative impact of climate change on wind power within the German Bight region.

In future work, we will consider a broader variety of data sources for bias correction, such as other wind atlas products, and wind farm power output data. Such data sources may be valuable in overcoming the relatively coarse resolution of the ERA-5 reanalysis data used within this work, as well as helping to overcome any biases in this data. We will also look to include further planned wind farms over larger spatial regions, and consider other regions worldwide where the relative magnitudes of climate and time-evolving wake effects may differ. This will enable the assessment of the impact of climate and wake effects on levelised cost of energy, and energy systems models.

%\FloatBarrier

\section*{Acknowledgements}

We acknowledge HPC resources and support from the Imperial College Research Computing Service (\url{http://doi.org/10.14469/hpc/2232}). The FINO1 data used within this study was made available by the FINO (Forschungsplattformen in Nord- und Ostsee) initiative, which was funded by the German Federal Ministry of Economic Affairs and Climate Action (BMWK) on the basis of a decision by the German Bundestag, organised by the Projekttraeger Juelich (PTJ) and coordinated by the German Federal Maritime and Hydrographic Agency (BSH).

%\bibliography{references}

\FloatBarrier
\appendix

\setcounter{figure}{0}
\setcounter{table}{0}

\section{Sensitivity to peak thrust coefficient}
\label{sec:sensitivity_to_ct}

As described in section \ref{sec:methods_study_region_and_farm_data}, the turbine power and thrust curves within this work were generated using PyWake's `generic\_wind\_turbines' module. While the power curve is typically well constrained by the turbine's physical parameters (rotor area, cut-in and cut-out speeds, etc), the thrust coefficient ($C_T$) curve can exhibit greater variation between turbine models. A key parameter in constructing `generic' $C_T$ curves is the peak $C_T$ value, which is assumed to be 0.8 by default. While this value is in reasonable agreement with publicly available turbine $C_T$ curves \citep{turbine_data}, in this section we investigate the influence of this parameter on the key results from this study. To do this, we repeated the experiments presented in section \ref{sec:results_build_out_effects} with peak $C_T$ values of 0.7 and 0.9. The results are summarised in table \ref{tab:sensitivity_to_ct}.

All wake effects increase in magnitude as the peak $C_T$ value increases. This is because the intensity and extent of turbine wakes increases with the turbine thrust. Larger peak $C_T$ values therefore induce greater wake losses. However, the overall trends are unchanged, and overall wake effects are still found to be of significantly greater magnitude than the mean climate impact, which has a value of -2.1\%. Nevertheless, these results demonstrate the significance of uncertainty due to unknown turbine data. Studies of wake losses would benefit from greater sharing of such data by turbine manufacturers. However, these results also highlight the uncertainty inherent in modelling future wind farm scenarios, where even the basic turbine parameters may be unknown.

\begin{table}[]
    \centering
    \begin{tabular}{c|c|c|c}
        Peak $C_T$ value & 0.7 & 0.8 & 0.9 \\ \hline
        Intra-farm wakes & -12.2\% & -13.3\% & -13.8\% \\
        Intra-cluster wakes & -4.0\% & -4.3\% & -4.4\% \\
        Inter-cluster wakes & -2.3\% & -2.5\% & -2.6\% \\ \hline
        Total wake effect & -18.4\% & -20.1\% & -20.8\% \\
    \end{tabular}
    \caption{Climate and wake effects for different values of the peak thrust coefficient ($C_T$), for the final build-out stage.}
    \label{tab:sensitivity_to_ct}
\end{table}

\setcounter{figure}{0}
\setcounter{table}{0}

\section{Alternative climate scenarios}
\label{sec:alt_climates}

In this section we present results from analysing the RCP2.6 scenario, in contrast with the RCP8.5 scenario which is the focus of the main text. We also present results from analysing an intermediate future period of 2040--2070, compared with the period of 2070--2100 which is assessed in the main text.

While the RCP8.5 scenario is a high-emissions scenario sometimes referred to as `business as usual', RCP2.6 is the lowest-emission scenario considered within CMIP5 (and hence EURO-CORDEX). Of the seven EURO-CORDEX models utilised to study the RCP8.5 scenario, data for the RCP2.6 scenario is available from only three. These are CNRM-A, NCC-R and MPI-R. For these three models, we repeated the experiments of section \ref{sec:results_climate_power} using RCP2.6 data. For each model and climate scenario, table \ref{tab:rcp_comparison} summarises the projected change in mean power for the 2027 build-out stage, between the future (2070--2100) and historical (1976--2006) periods.

For the CNRM-A model, the projected impact of climate change on wind power is significantly greater for the RCP8.5 scenario than for the RCP2.5 scenario. However, for the NCC-R and MPI-R models, the impact is greater under the RCP2.6 scenario than RCP8.5. Averaging across the three models, there is a small increase in the climate impact between RCP2.6 and RCP8.5.

Increased impact from lower-emission scenarios is not necessarily surprising, given the complexity of the earth system and its (modelled) response to human activity. In particular, we note also that this study focuses on only a single location, hence the results will be sensitive to climate-induced changes in spatial patterns. This is also consistent with previous studies. For example, \citet{fernandez2023dynamic} downscaled projections from the CESM2 model, with their results showing differing signs of change in wind power density over the North Sea, depending on the climate scenario.

\begin{table}[]
    \centering
    \begin{tabular}{c|c|c}
         & RCP2.6 & RCP8.5 \\ \hline
        CNRM-A & -2.3\% & -4.2\% \\
        NCC-R & -1.1\% & -0.5\% \\
        MPI-R & -1.0\% & -0.1\% \\ \hline
        Mean & -1.4 & -1.6\% \\
    \end{tabular}
    \caption{Comparison of climate change effects for the RCP2.6 and RCP8.5 scenarios. Only models for which both climate scenarios are available are shown.}
    \label{tab:rcp_comparison}
\end{table}

Table \ref{tab:period_comparison} compares projected changes in wind power output for the 2027 build-out, for the period 2040--2070, and the period 2070--2100. We again find that there is a lack of agreement between climate projections as to which future period experiences the greatest decrease in wind power compared with the historical period. The mean effects are similar, although the 2070--2100 period is projected to produce slightly lower wind power than the 2040--2070 period. These results are again not unexpected; for example, \citet{martinez2023evolution} assessed changes in wind power density for Northern Europe including the North Sea, based on CMIP6 projections. Their results showed that the effect of climate change on wind power density is not necessarily monotonic with time.

\begin{table}[]
    \centering
    \begin{tabular}{c|c|c}
         & 2040--2070 & 2070--2100 \\ \hline
        CNRM-A & -2.9\% & -4.2\% \\
        NCC-A & -2.1\% & -2.2\% \\
        MOHC-A & -2.5\% & -3.8\% \\
        CNRM-R & -0.3\% & -0.9\% \\
        NCC-R & -1.6\% & -0.5\% \\
        ICHEC-R & -2.0\% & -2.9\% \\
        MPI-R & -2.2\% & -0.1\% \\ \hline
        Mean & -1.9\% & -2.1\% \\
    \end{tabular}
    \caption{Comparison of climate change effects for the RCP8.5 scenario, for the periods 2040--2070, and 2070--2100. For both time periods, the changes are calculated relative to the historical period.}
    \label{tab:period_comparison}
\end{table}

Overall, the results of this section highlight the inherent uncertainty in climate projections, and the need for model ensembles in quantifying this uncertainty.

\end{document}